\RequirePackage{snapshot}	

\documentclass[final,5p,times,twocolumn]{elsarticle} 

\usepackage{url}
\usepackage{lineno,hyperref}
\usepackage{amssymb}
\usepackage{subfigure}
\usepackage{placeins}
\usepackage{mathtools}
\usepackage{amsmath}
\usepackage{gensymb}
\usepackage{svg}
\journal{Nuclear Instruments and Methods in Physics Research A}

\usepackage[
, parse-units=false		
, uncertainty-mode = separate	
, range-phrase= - 		
, range-units = single 		
, free-standing-units=true, space-before-unit=true, use-xspace=true
, quantity-product = \,		
]{siunitx}%
\usepackage{chemmacros}	

\usepackage{graphicx}
\usepackage{caption}
\usepackage{subcaption}
\graphicspath{{.}{Figures}}
\newcommand*{\sect}[1]{Section~\ref{sect:#1}}
\newcommand*{\tab}[1]{\tablename~\ref{tab:#1}}
\newcommand*{\fig}[1]{\figurename~\ref{fig:#1}}

\newcommand*{\xbar}[1]{
  {}\mkern3mu\overline{\mkern-3mu#1\mkern-1mu}\mkern1mu}

\begin{document}
\begin{frontmatter}

  \title{Development and Testing of a Modular Large-Area Cosmic Ray
    Telescope Using Scintillator-Fiber Hybrid Design for
    Millimeter-Level Muon Tracking}

  \author{Yan Niu}%
  
  \author{Anqing Wang} \author{Xiangxiang Ren} \author{Dong Liu
  }
  \author{Meng Wang }

  \affiliation{
    addressline={Research Center for Particle Science and Technology},
    organization={Shandong University}, city={Qingdao},
    postcode={266237}, country={China}}

  \begin{abstract}
    Cosmic-ray muons, owing to their high penetration power and
    abundance, have been widely employed as a natural probe in
    experimental particle physics.
    We developed a meter-scale cosmic-ray muon telescope, consisting
    of two parallel super-layers (\SI{1}{m} $\times$ \SI{1}{m})
    separated vertically by one meter.
    A super-layer is composed of two orthogonal detection layers,
    each one consisting
    of eighteen modules arranged in parallel and packed closely
    together. A module consists of a plastic scintillating bar
    precisely aligned and stacked on top of an underlying
    scintillating fiber mat
    in which fibers are arranged in a row of bundles. For a detection
    layer, each scintillator bar is coupled to a PMT while fiber
    bundles at the same position within all modules are coupled to a
    single PMT. Signals from scintillating bars and fibers are
    combined together to determine hit positions. With this detection
    scheme, the telescope can
    meet the requirement of spatial resolution and
    reduce the number of readout electronic channels.

    This article presents the comprehensive development of the
    telescope, encompassing its geometric design, data acquisition
    system, and performance evaluation.
    Experimental results show that the telescope achieves a position
    resolution better than \SI{2}{mm} and an overall detection
    efficiency of $\sim$85\%.
    The innovative design keeps the manufacturing cost low while
    maintaining high spatial resolution and detection efficiency.
  \end{abstract}


\end{frontmatter}


\section{Introduction}
\label{sect:intro}
The High Energy cosmic-Radiation Detection (HERD) facility is a future
space experiment to be operated 
on the Chinese Space Station, with an expected in-orbit lifetime of at
least ten years~\cite{Betti:2025zuu, Mori:2022jyj}.
The core sub-detector of HERD is a three-dimensional imaging
calorimeter (CALO), which consists of 7,500 LYSO cubic crystals
measuring 3 $\times$3 $\times$ \SI{3}{cm^{3}} and forms an
approximately spherical geometry~\cite{HERD:2021ufz,HERD:2023awk}.
One ground calibration method of the CALO relies on cosmic-ray muons,
demanding a telescope with millimeter-level spatial resolution.

A meter-scale cosmic-ray muon telescope has been developed for this
purpose. The telescope reduces the number of readout electronic
channels by matching the signals from plastic scintillating bars and
fibers. This innovative design ensures the required detection
precision and effectively decreases the manufacturing cost.
Iterative analyses of Monte Carlo simulations and prototype tests were
conducted during the design process.

Muon imaging has demonstrated broad application potential in mineral
exploration~\cite{Malmqvist:1979,Schouten:2018kla,Liu:2024pxy},
archaeological void
detection~\cite{Alvarez:1970ecc,Saracino:2017,Morishima:2017ghw}, and
customs inspection~\cite{Schultz:2004kx,Hogan:2004rn}, with the
capability to maintain effective detection performance under complex
environmental conditions.
More details can be found in a review~\cite{Bonomi:2020dmm}.
For cosmic muon applications using scintillating bars or fibers, the
spatial resolution is proportional to the scintillator lateral size
($x$), while the number of electronic channels is in general inversely
proportional to $x$.
For instance, an application using scintillating bars of a triangle
shape ($x=33\mm$) reaches the spatial resolution of about 3\mm, and
has 32 channels per detection layer~\cite{DErrico:2022ffa,
  Anastasio:2013ela}.
In our approach, the number of channels is inversely proportional to
the \emph{square root} of $x$ (see \sect{geometric}),
and we have the spatial resolution
better than 2\mm with 36 channels per detection layer.
Even a sub-millimeter resolution, in principle, can be reached with
scintillating fibers of 1\mm diameter without increasing the number of
channels significantly.
Hence our method in reducing the number of electronic channels
provides an economical solution for cosmic muon applications demanding
high spatial resolution.

\sect{Design} details the geometric design of the telescope and its
circuit logic, including key technologies for data acquisition (DAQ)
and processing.
\sect{Verification} presents the systematic performance evaluation of
the detector components.
\sect{Resolution and efficiency} showcases the performance test
results, which are critical for assessing its detection capabilities,
with a focus on core metrics such as spatial resolution and detection
efficiency.
\sect{Summary} summarizes the main achievements of this study.

\section{The Meter-Scale Cosmic-Ray Muon Telescope}\label{sect:Design}

\begin{figure*}[!htb] %
  \centering%
  \hspace{\stretch{1}}%
  \includegraphics[height=0.21\textheight]{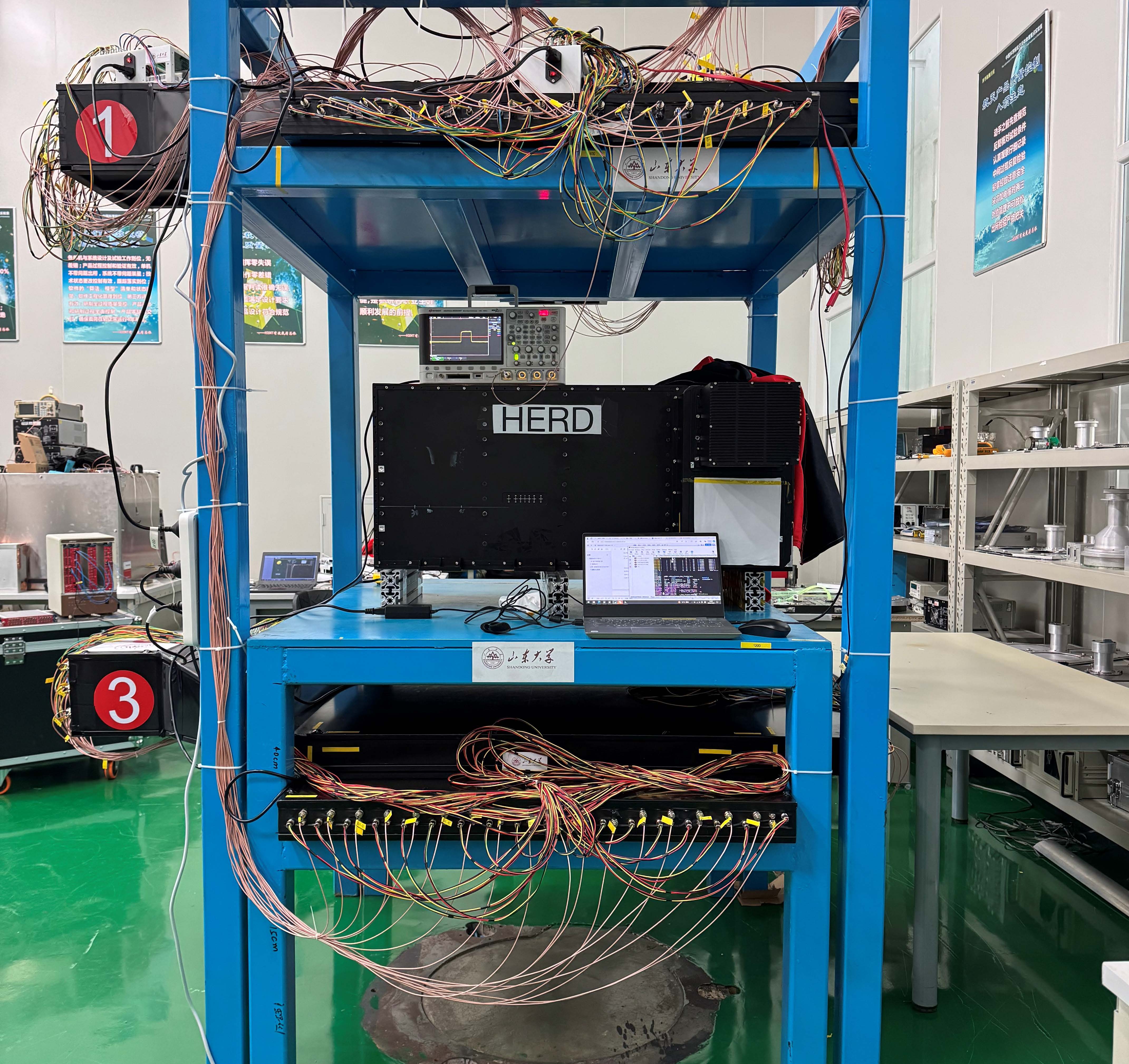}%
  \hspace{\stretch{1}}%
  \includegraphics[height=0.21\textheight]{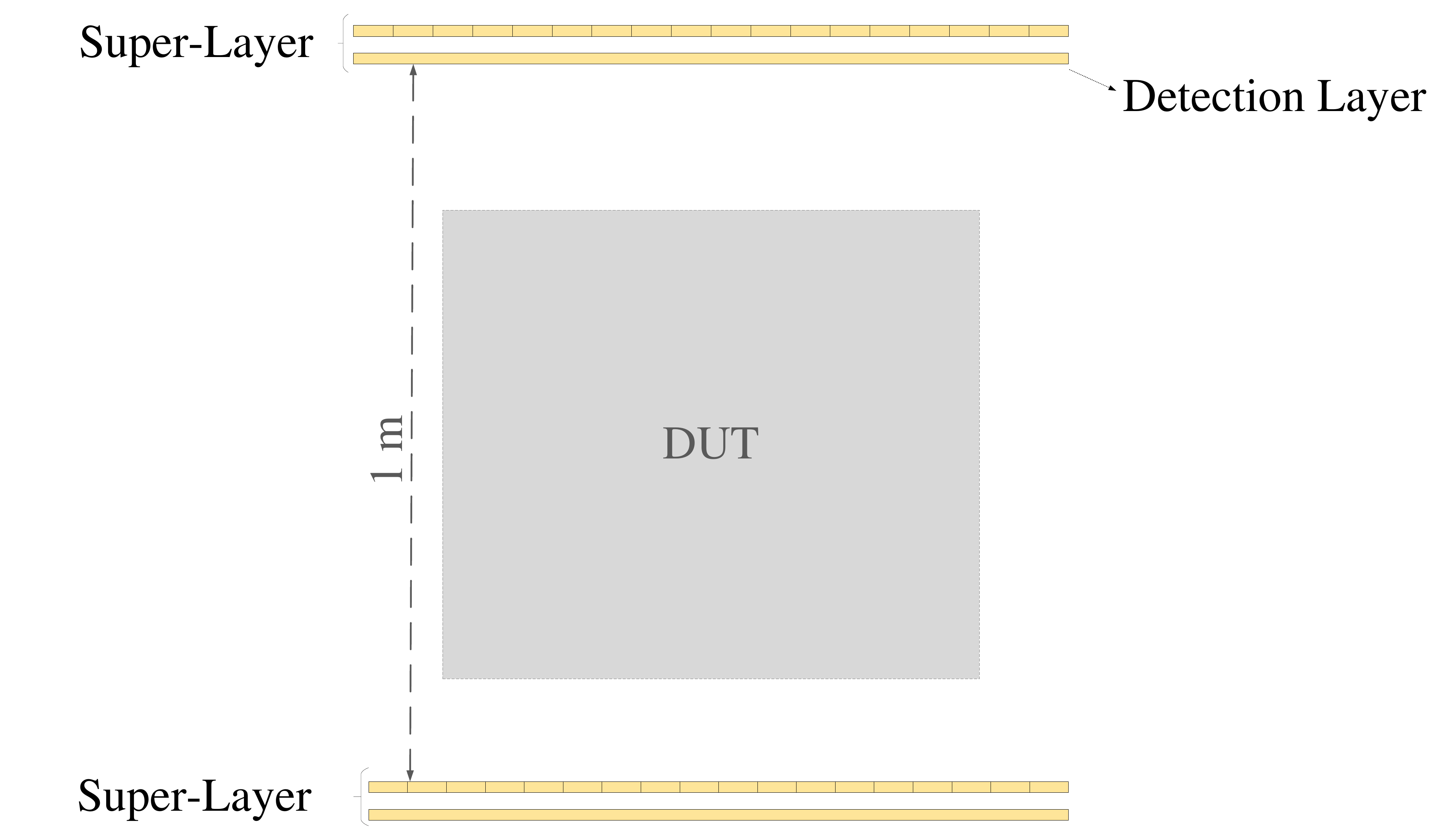}%
  \hspace{\stretch{1}}
  \par
  \caption{Photo (left) and schematic diagram (right) of the telescope
    test setup.}
  \label{fig:HerdFourLayer}
\end{figure*}

\subsection{Geometric Structure Design}\label{sect:geometric}
\begin{figure}[!htb]
  \centering%
  \includegraphics[width=0.82\linewidth]{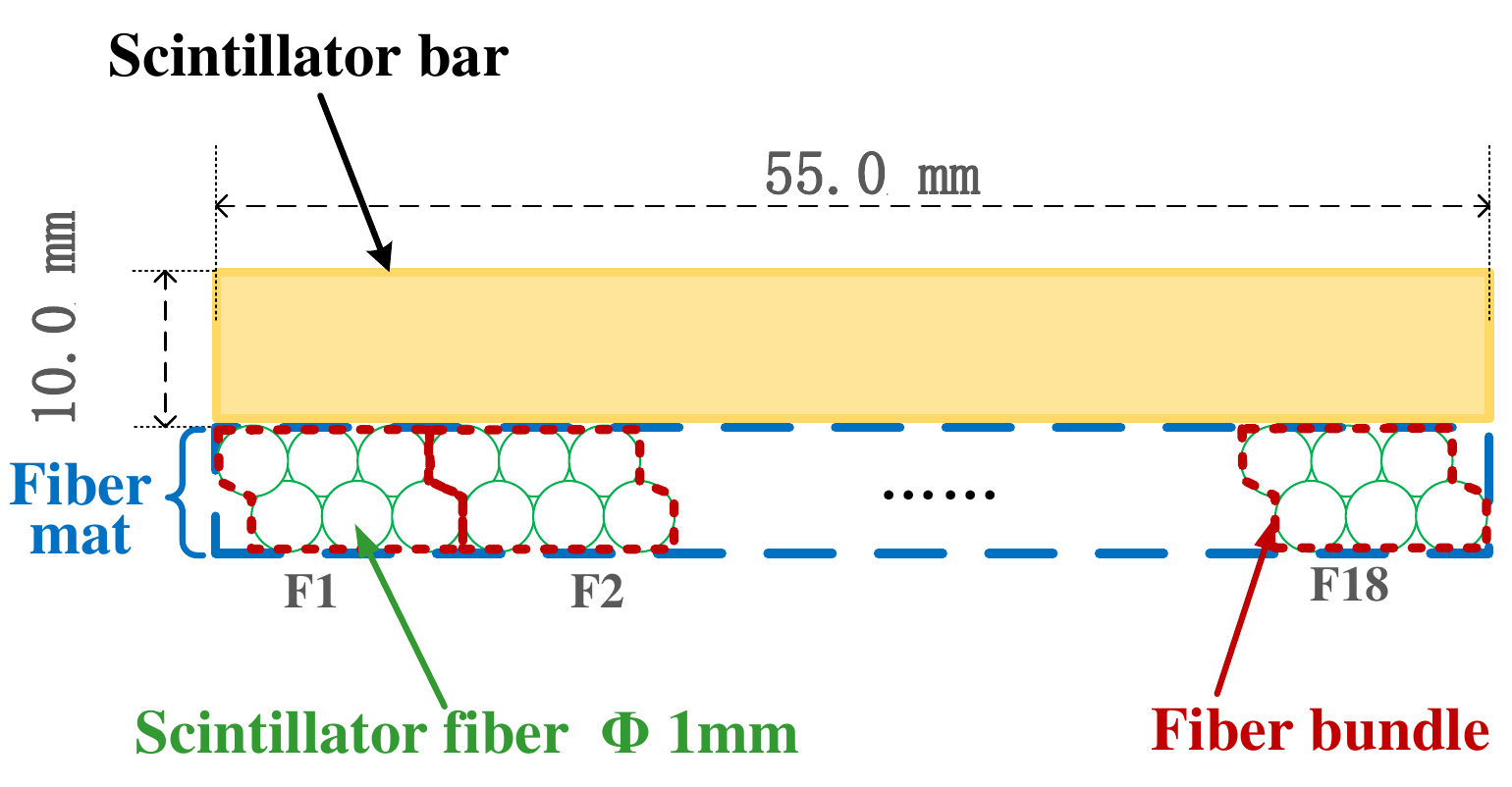}
  \caption{Cross-sectional structure of a module.}
  \label{fig:Cross_section}
\end{figure}

The telescope consists of two \emph{super-layers}, each about $1\metre
\times 1\meter$ in area, parallel to each other
and vertically separated by about \SI{1}{\meter}, as shown in
\fig{HerdFourLayer}.
Each {super-layer} comprises two functionally identical
\emph{detection layers} arranged orthogonally.
A single {detection layer} comprises eighteen \emph{modules}, as shown
in \fig{Cross_section}, arranged side by side in a close-packed
configuration.
Each {module} consists of a plastic \emph{scintillating bar} and a
\emph{scintillating fiber} mat.
%
The scintillating bar, each coupled to a photomultiplier tube (PMT),
provides the trigger signal and coarse positioning.
%
%
Fibers in the mat are grouped in bundles (indicated as F1, F2, ... in
\fig{Cross_section}). Each fiber bundle is sent to a different PMT for
precise position determination. However, to reduce the number of
readout channels, fiber bundles at the same relative position in the
18 modules are connected to the same PMT.

The scintillating bar is a Haotang-Xinghe SP101 plastic scintillator
with the dimension of
$\SI{55}{\milli\meter} \times \SI{10}{\milli\meter} \times
\SI{1000}{\milli\meter}$.
All surfaces except at the end ports, are coated with a titanium
dioxide \ch{TiO2}-mixed resin. The \ch{TiO2} coating enhances the
internal reflection of scintillation light produced by ionizing
radiation.

The fiber mat consists of two layers of Kuraray
SCSF-78~\cite{KuraraySCSF78} scintillating fibers, each
\SI{1}{\milli\meter} in diameter.  Three fibers from the upper layer
and three fibers from the lower layer are grouped into a single
bundle.
Hence, the \emph{fiber bundle} has a width of about
\SI{3}{\milli\meter}.
Simple calculations indicate that, for a detection layer, the number
of readout channels ($N$) is minimized when the number of modules
($M$) matches the number of fiber bundles in a module ($F$),
\begin{equation*}
N = M + F = \left(\sqrt{M} - \sqrt{F} \right)^2 + 2\sqrt{M\cdot F}
  \ge 2\sqrt{M\cdot F}.
\end{equation*}
To meet the requirement of about $\SI{1}{mm}$ spatial resolution, the
width of the fiber bundle ($x$) is determined to be,
$x \approx \sqrt{12} \cdot \SI{1}{mm} = \SI{3}{mm} $.
For a meter-scale detector, the number of modules is determined to be,
$M \approx \sqrt{\SI{1}{m}/x} = 18$, and the width of a scintillator
bar, $w \approx Mx = \SI{54}{mm}$, is chosen to be \SI{55}{mm}.

\begin{figure}[!htb]
  \centering%
  \includegraphics[width=0.35\textwidth]{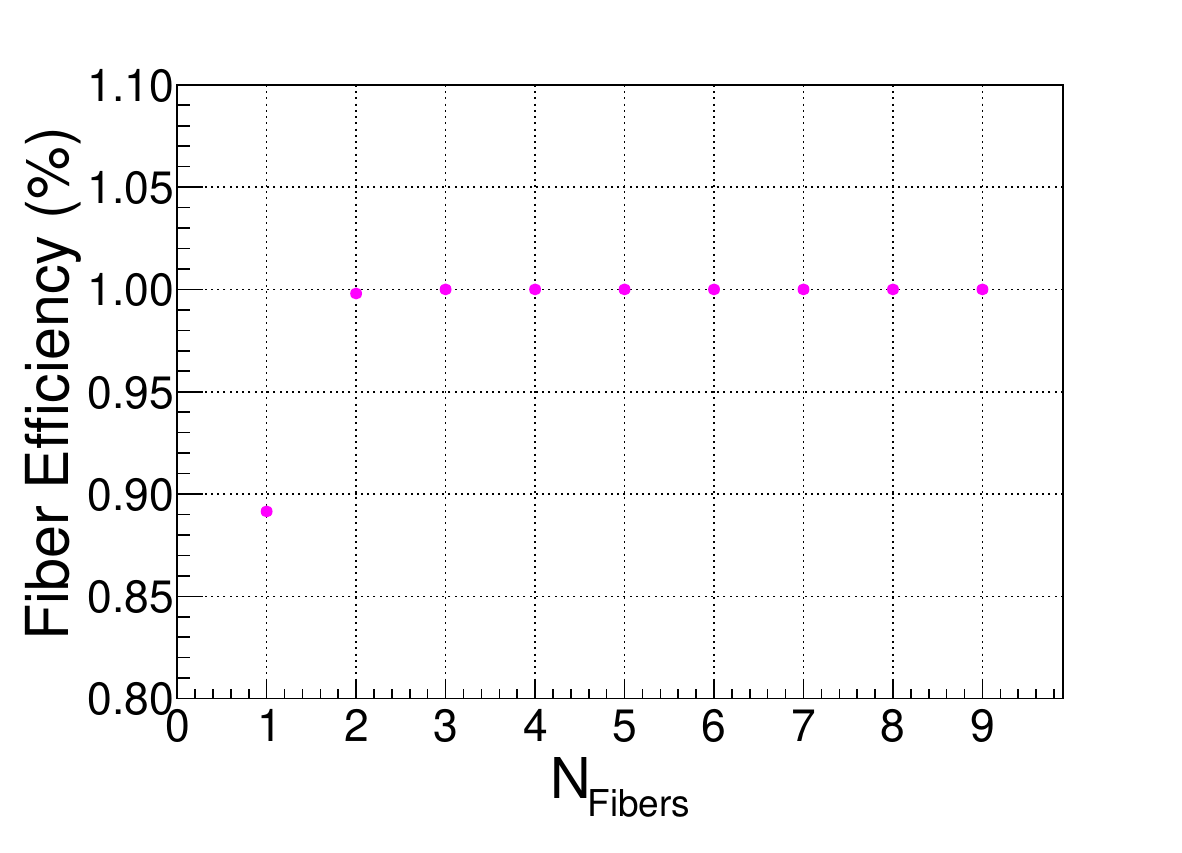}
  \caption{Detection efficiency of scintillating fiber mat as a
    function of the number of fiber layers in the mat
    ($N_\text{Fibers}$). The distribution is obtained with the Geant4
    simulation.}
  \label{fig:GeantNfiber}
\end{figure}

The number of fiber layers in the mats was determined through detailed
Geant4 simulations~\cite{GEANT4:2002zbu, Allison:2006ve}.  The results
indicate that a double-layer fiber configuration achieves a detection
efficiency of 99\%, as shown in \fig{GeantNfiber}.

\begin{figure}[!htb]
  \centering%
  \includegraphics[height=28mm]{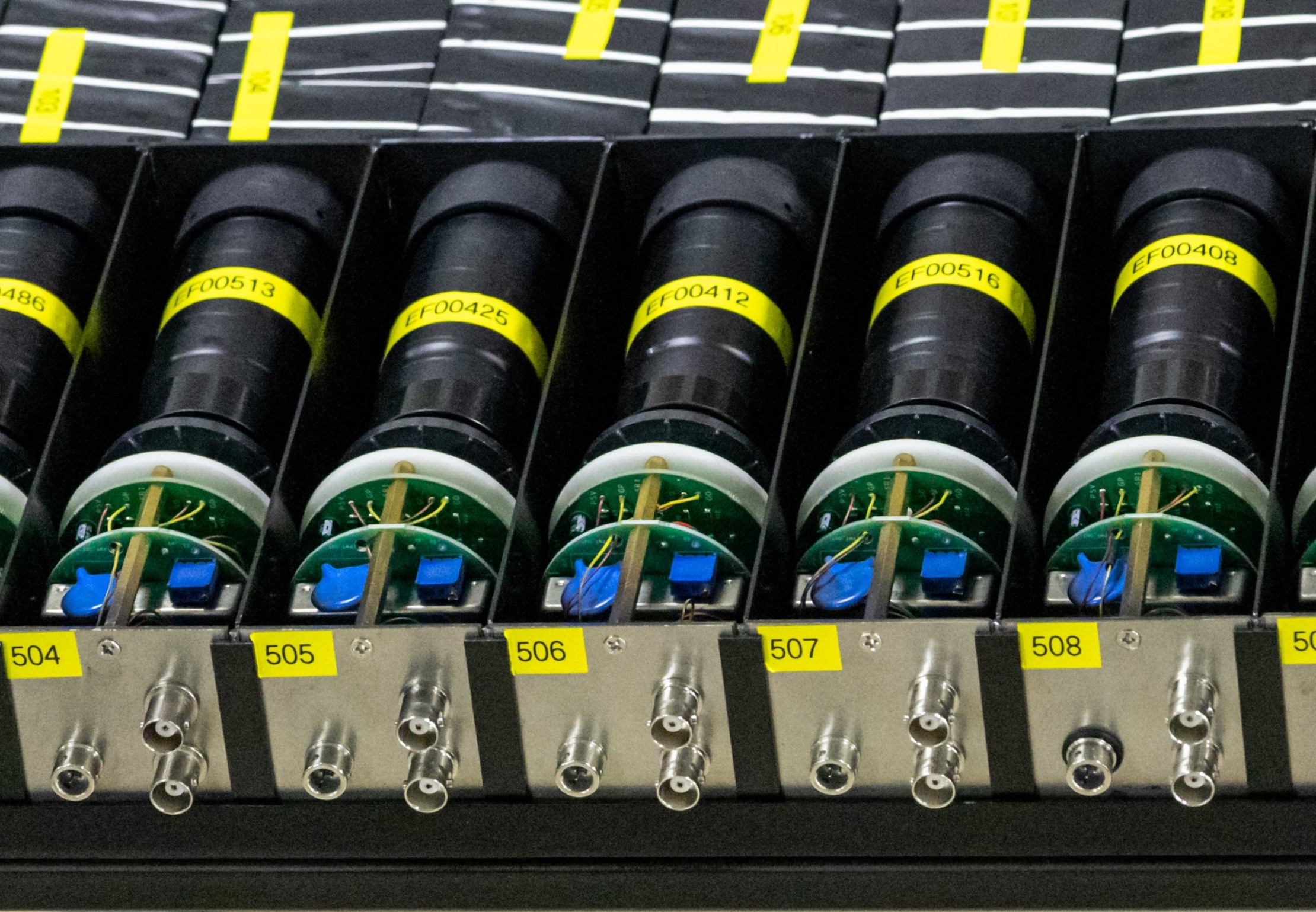}~%
  \includegraphics[height=28mm]{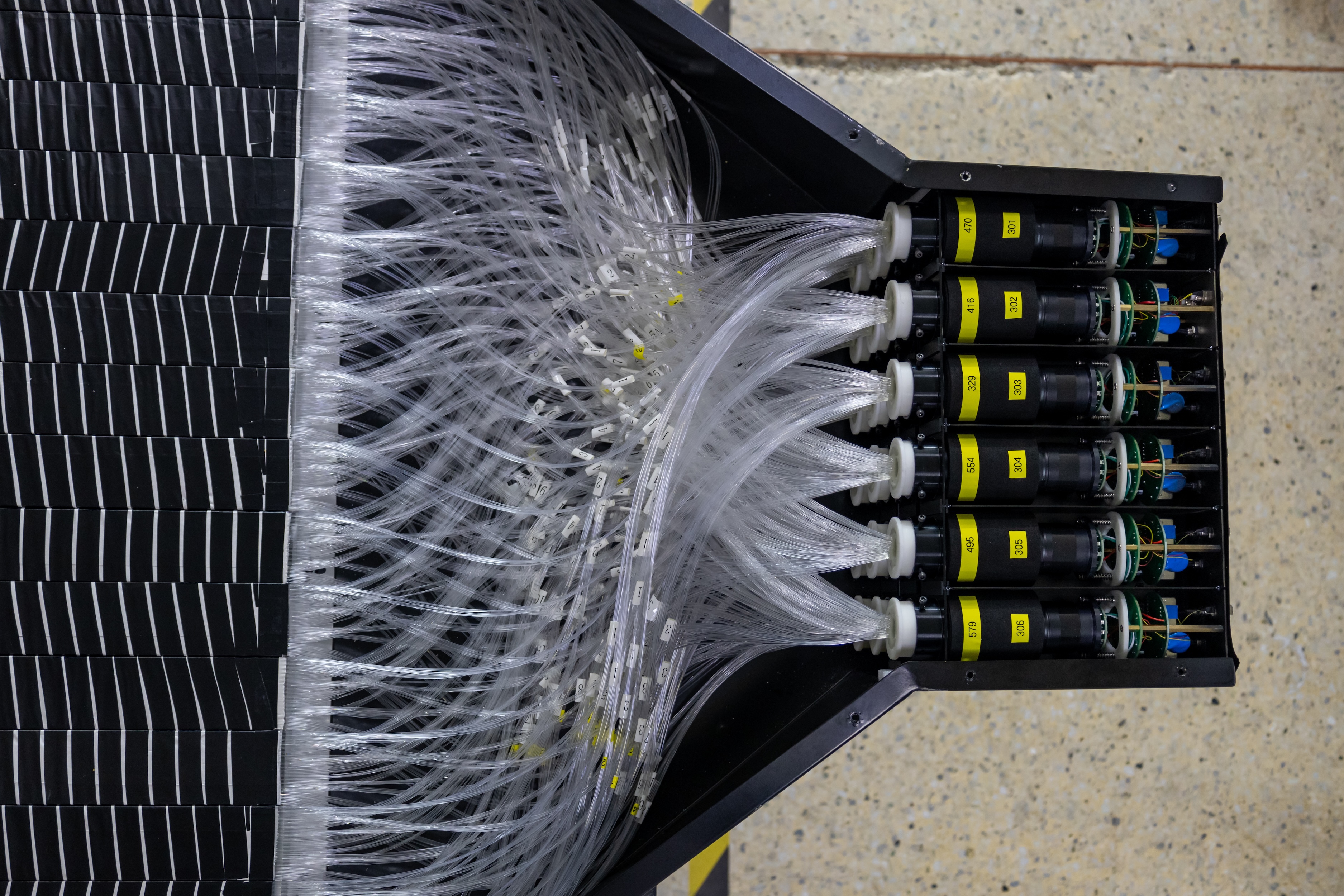}
  \caption{PMT coupling with scintillating bars (left) and fibers
    (right).}
  \label{fig:PMTbase}
\end{figure}

\begin{figure}[!htb]
  \centering%
  \includegraphics[width=0.45\linewidth]{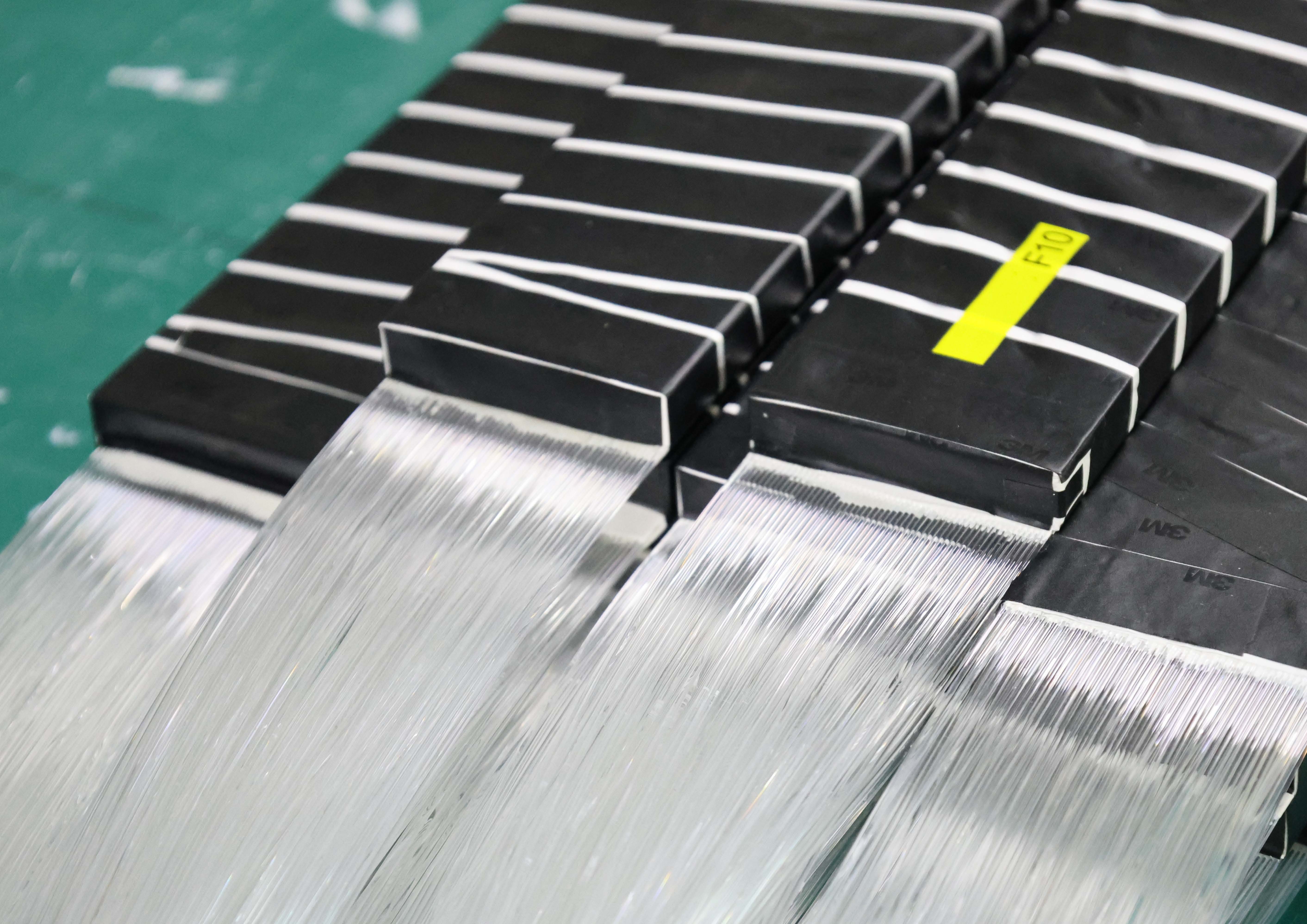}~%
  \includegraphics[width=0.45\linewidth]{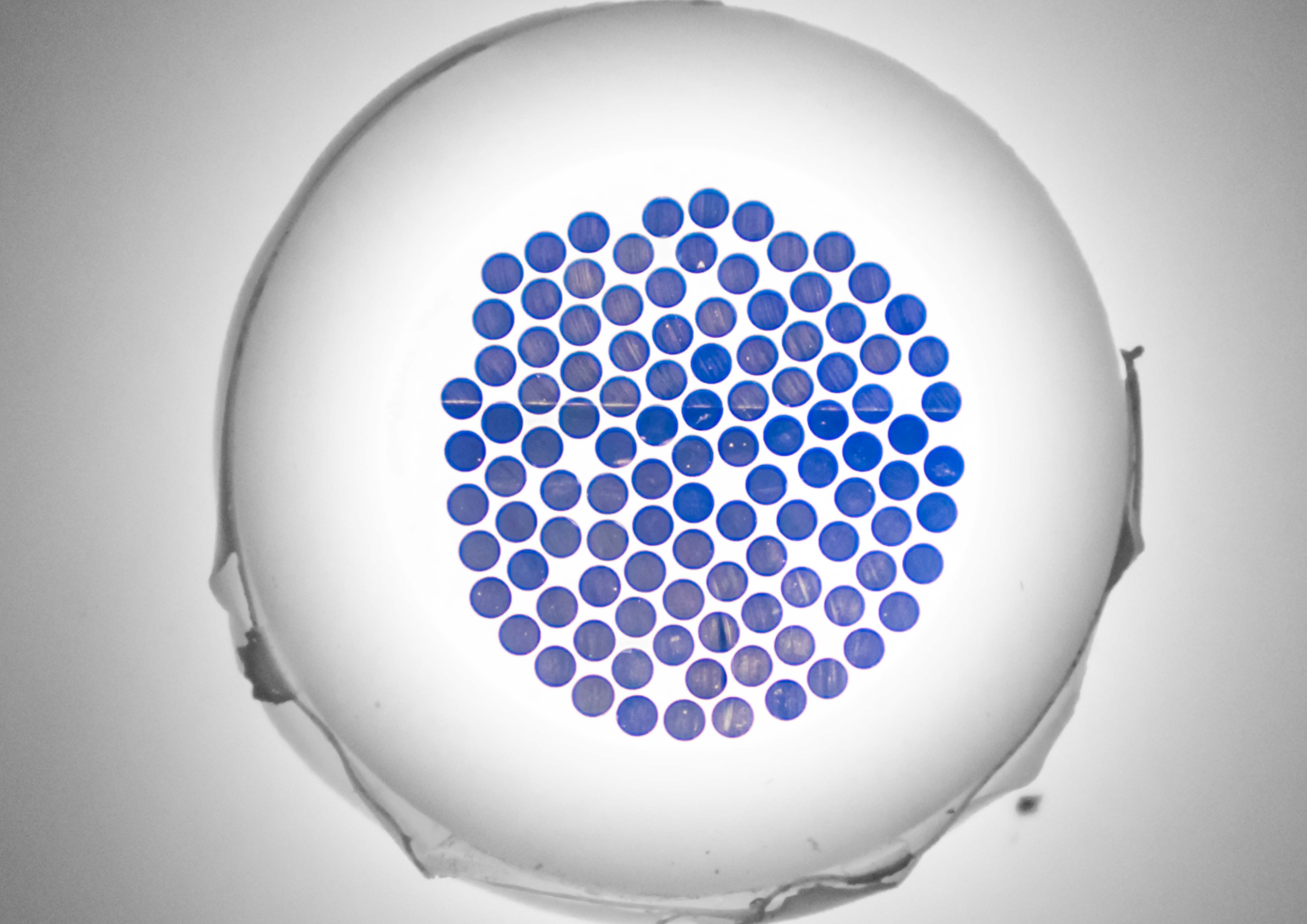}
  \caption{The assembled modules (left) and polished end-face of
    scintillating fibers (right), which to be coupled to a PMT. }
  \label{fig:ProductModule}
\end{figure}

Hamamatsu CR285
PMTs~\cite{BeijingHamamatsuCR285} are utilized to detect
scintillation light.
As shown in \fig{PMTbase}, PMTs are directly coupled to
scintillating bars one by one (left), whereas each PMT is coupled to
108 ($6\times18$) fibers, one bundle from each module (right).
Glass envelopes of all PMTs are wrapped with black insulating tape and
covered with black foam to ensure complete light shielding and
protection.
The end face of fiber bundles intended for coupling with the PMT
underwent precise polishing to achieve optical flatness, as shown in
\fig{ProductModule} (right). A layer of optical grease was uniformly
applied between the fiber end face and the PMT photocathode window to
improve photon transmission efficiency.
The other end of fibers in a mat was sealed with a reflective coating
to enhance the fibers' response.
The scintillating bar–PMT interface follows the same protocol of
polishing and optical grease.

\begin{figure}[!htb]
  \centering%
  \includegraphics[width=0.82\linewidth]{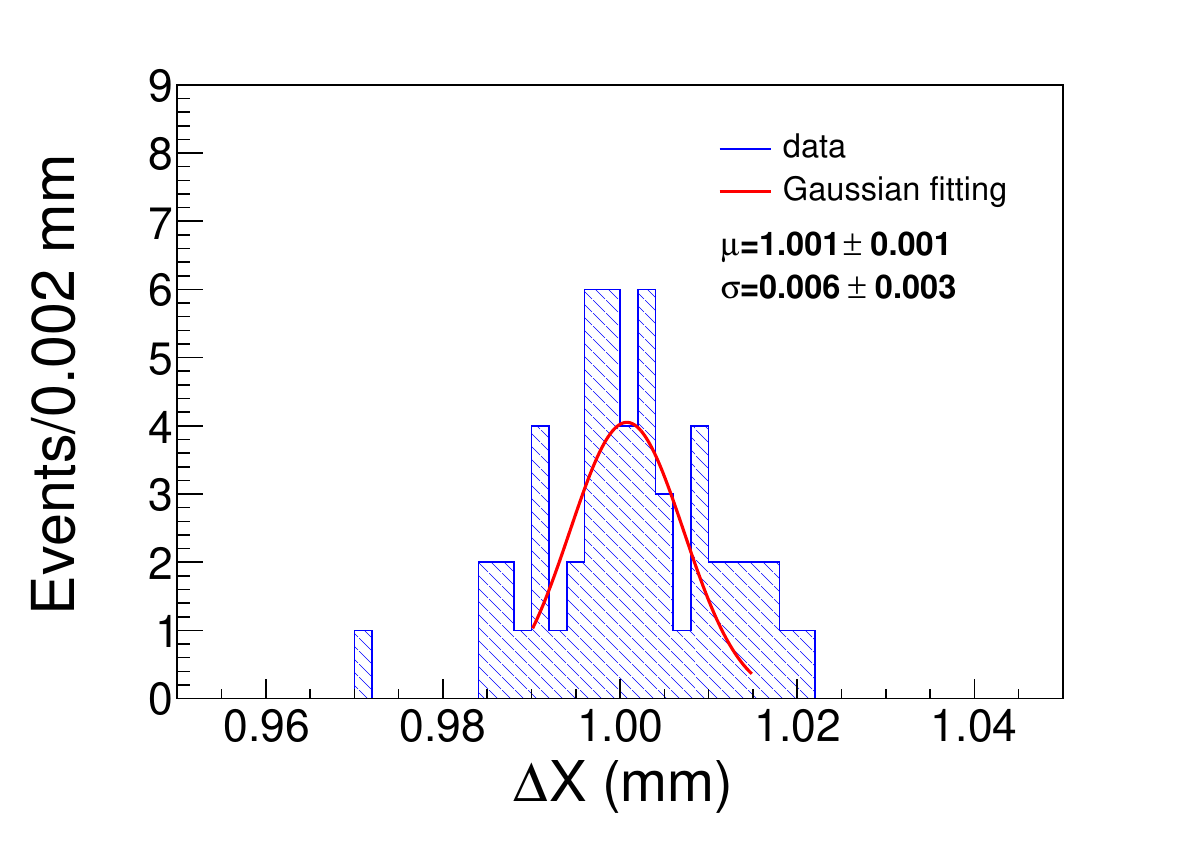} %
  \caption{Distribution of spacing between adjacent fiber centers in a
    module.}
  \label{fig:ProductFiber}
\end{figure}

The fiber mat fabrication process started by aligning one end of
fifty-four fibers,
each one more than \SI{2}{\meter}-long
. The initial
one-meter segment of the aligned fibers was placed into fifty-four
parallel grooves of a dedicated fixture and secured with clamps on
both sides.
The entire clamped section was then coated with \ch{TiO2}-doped epoxy,
pressed firmly with a metal bar, and cured to form a tightly packed
and stable fiber mat consisting of fifty-four fibers.
The remaining length of each fiber extended freely from the opposite
end of the fixture, forming a neat bundle.
The central positions of the fibers in the manufactured fiber mat were
measured
with a length measuring instrument with the resolution of 5\um,
and the spacing between adjacent fibers are shown in
\fig{ProductFiber}.  The measurements confirm that the fibers in the
mat are densely packed, with a positional deviation within
\SI{0.01}{mm}.
The fabricated scintillating bar and the cured end of the fiber mat
were aligned and bonded using \ch{TiO2}-doped epoxy, with a stacking
length of 1 meter. The entire stacking section was wrapped and sealed
with 1056 tyvek~\cite{Arteaga-Velazquez:2005arc}, followed by tight
winding with black insulating tape to ensure both mechanical stability
and optical isolation of the module, as shown in
\fig{ProductModule} (left).

\subsection{Operating Principle}

\begin{figure*}[!htb]
  \centering%
  \includegraphics[width=0.72\textwidth]{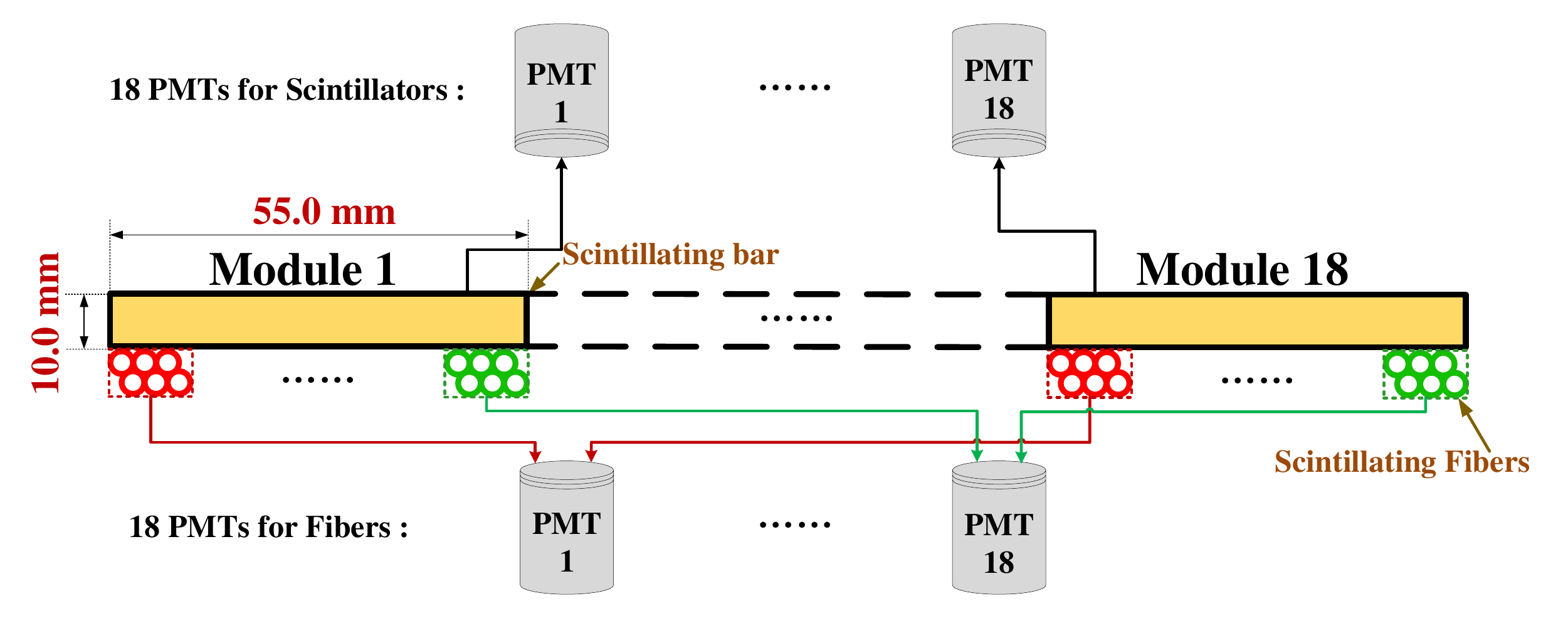}
  \caption{Working principle of position locating in a single
    detection layer.  Each scintillating bar (yelllow rectangles) is
    coupled to a PMT, while fiber bundles (circles in dashed boxes)
    with the same numbering (indicated by color) from all 18 modules,
    are combined and coupled to a PMT.}
  \label{fig:HerdSingleLayer}
\end{figure*}

The working principle of position locating in a single detection layer
is shown in \fig{HerdSingleLayer}.
Each detection layer consists of eighteen modules arranged in
parallel.
The fiber mat in a module is composed of eighteen bundles, which are
numbered sequentially from left to right.
Fiber bundles with the same number from all 18 modules are combined
and coupled to a single PMT, as shown in \fig{PMTbase} (right).
Each scintillating bar, from the other end relative to that of fibers,
is individually coupled to a PMT.

A cosmic-ray muon passing throught a detection layer gives signals in
both a scintillating bar and a fiber bundle. By matching the two
signals, the millimeter presicion positioning can be obtained.
Position reconstruction within a \SI{1}{m} $\times$ \SI{1}{m} area is
achieved using 36 PMTs per detection layer. The total number of PMT
readout channels for the entire telescope amounts to 144.

\subsection{Data Acquisition (DAQ) System}
\label{sect:schematic}
\begin{figure*}[!htb] 
  \centering%
  \includegraphics[width=0.71\linewidth]{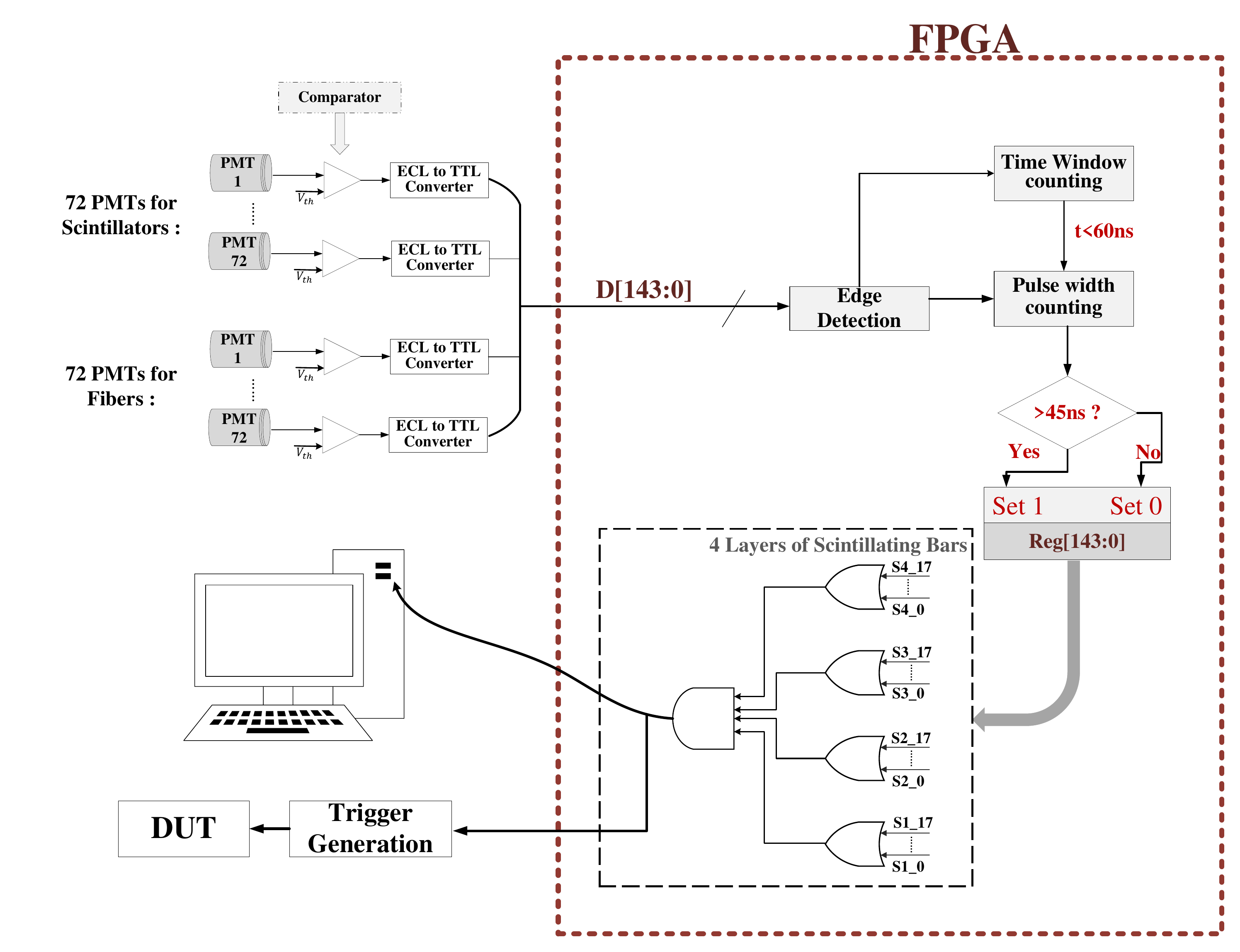}
  \caption{Schematic diagram of the telescope DAQ system, see text for
    details.
  }
  \label{fig:LogicCircuit}
\end{figure*}

\begin{figure}[!htb]
  \centering%
  \includegraphics[width=0.62\linewidth]{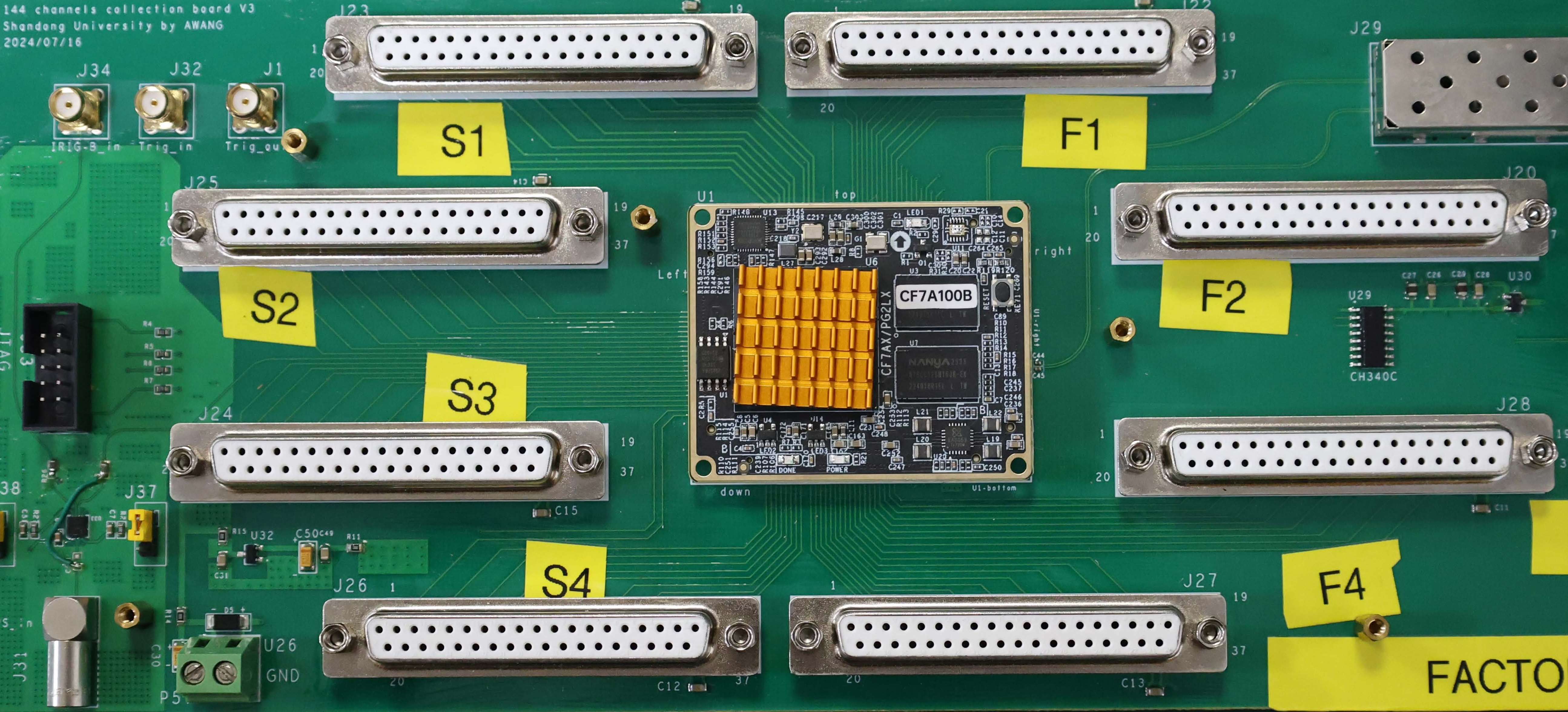}
  \caption{The FPGA board used in the telescope.}
  \label{fig:Circuit}
\end{figure}

The DAQ system mainly consists of three parts: signal collection,
signal processing and data storage, as shown in
\fig{LogicCircuit}.
The negative pulses from 144 PMTs are digitized in parallel via
time-over-threshold (TOT) circuits, with individual thresholds set for
each channel.
The TOT signals are then processed by a Field Programmable Gate Array
(FPGA) board, as shown in \fig{Circuit}, to find cosmic-ray muons
passing through the whole telescope.
Details of the signal processing in FPGA are described as the
following.
\begin{itemize}
\item Upon the arrival of the first TOT signal, a coincidence time
  window (\SI{60}{ns}) is open.
  The coincidence window ensures temporal correlation among signals
  from different detection layers.
  
\item A valid \emph{hit} signal is required that the TOT signal
  elapses longer than a fixed time width (\SI{45}{ns}) within the
  coincidence window. The validation results are saved in register
  bits, 1 for hit and 0 for none.

\item After the coincidence window, register bits corresponding to
  scintillator bars are checked, and a \emph{trigger} is issued when there
  are at least one hit for each detection layer, otherwise the
  register is reset.
  
\item The trigger initiates transferring all register bits to a PC for
  recording; meanwhile, it is also sent to the detector under test
  (DUT).
  The transmission rate can be \SI{1.5}{Mbps} (USB UART) or
  \SI{1000}{Mbps} (Ethernet TCP), upon request.
\end{itemize}

All relevant DAQ parameters, e.g. thresholds and time windows, are
configurable according to application requirements.
The mentioned values above are used in the calibration of the HERD
CALO.

\section{Performance Verification of Detector Components}
\label{sect:Verification}

\subsection{PMT Characterization}
A PMT bench~\cite{Wang:2015cyx} was utilized to evaluate the gain of
each PMT, enabling efficient batch characterization.

The PMTs were placed in a light-tight testing enclosure. A pulsed LED
driven by a pulse generator delivered light to each PMT via a
\SI{1}{mm} diameter optical fiber. A multi-channel high-voltage power
supply provided bias voltage to all PMTs. The DAQ system for the PMT
characterization was built with VME and NIM standard modules. A
central workstation controlled all equipments, including the pulse
generator, power supply and the DAQ system.

\begin{figure}[!htb]
  \centering%
  \includegraphics[width=0.72\linewidth]{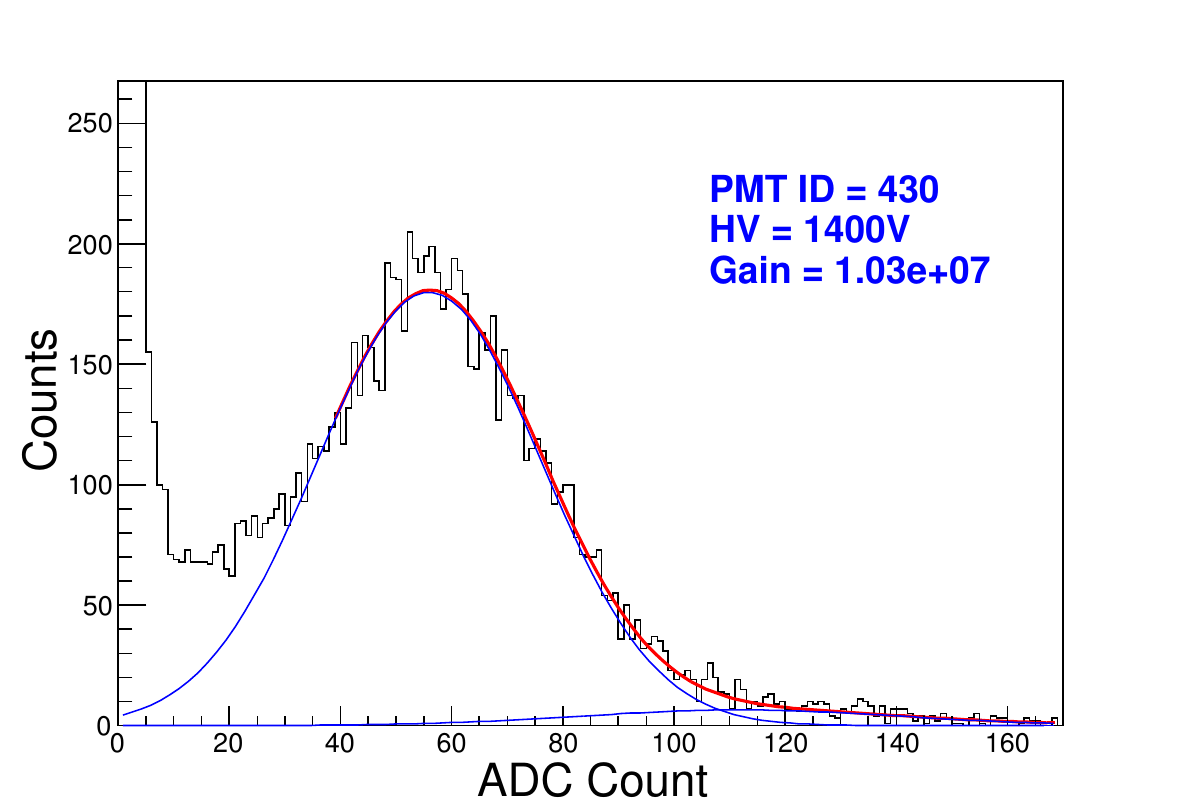}
  \caption{The charge spectrum of a PMT. The left peak is pedestal,
    and the right distribution is fitted (red) with two component
    Gaussians (blue) indicated. The dominant Gaussian is of SPE.  }
  \label{fig:SPE}
\end{figure}

\begin{figure}[!htb]
  \centering%
  \includegraphics[width=0.72\linewidth]{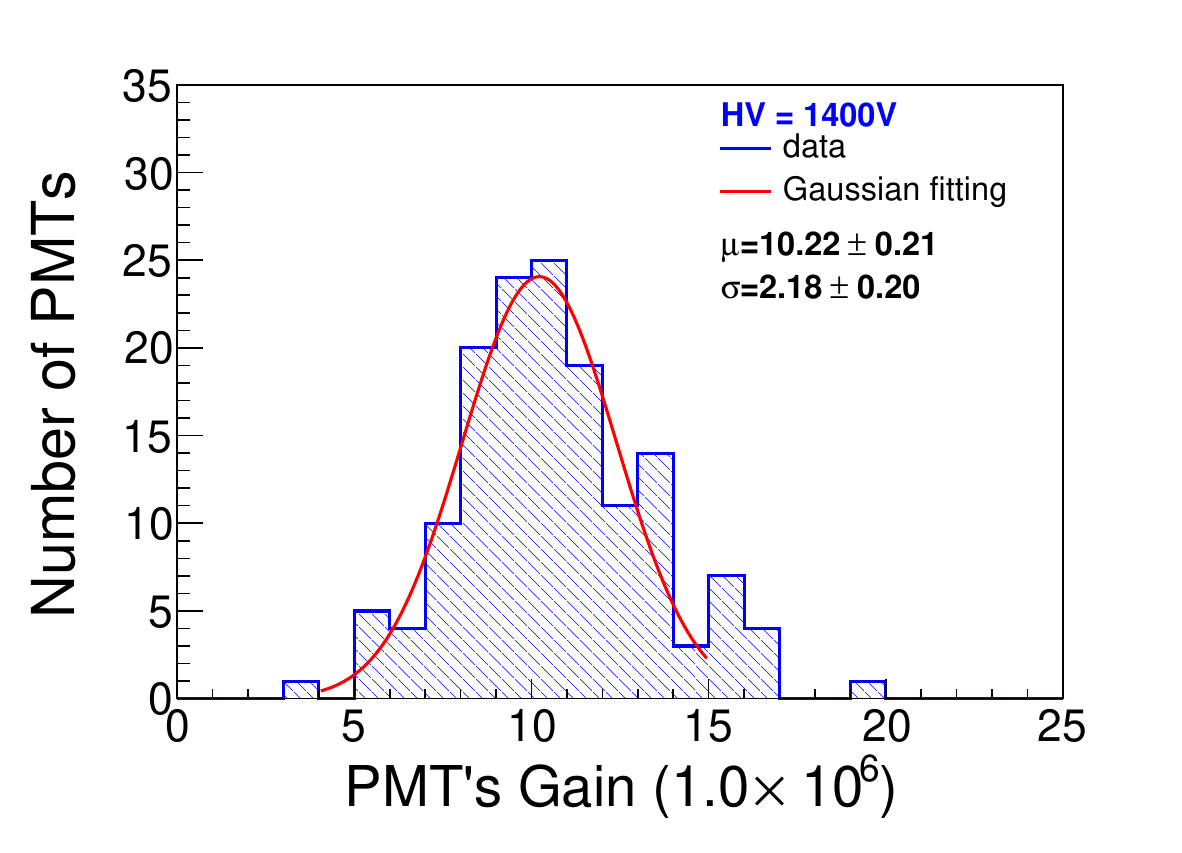}
  \caption{Distribution of gain for all PMTs operating at the high
    voltage of \SI{1400}{V}. The blue shaded region represents the
    measured data, and the red curve indicates the Gaussian fit to the
    distribution.}
  \label{fig:PMTGain}
\end{figure}

The PMT characterization was performed with the single photoelectron
(SPE) technique by
adjusting
the LED light intensity low enough to
ensure the SPE response of PMTs. An example of the SPE spectrum is
shown in \fig{SPE}.
The gain distribution measured at \SI{1400}{V} follows a Gaussian
profile with $(\mu, \sigma)=(12.22, 2.18) \times 10^{6}$, demonstrating
stable performance across all PMTs and confirming their
suitability for the telescope Assembly, as shown in \fig{PMTGain}.

\subsection{Single Module Test}
\label{sect:Single_Module_Testing}
\begin{figure}[!htb]
  \centering%
  \includegraphics[width=0.62\linewidth]{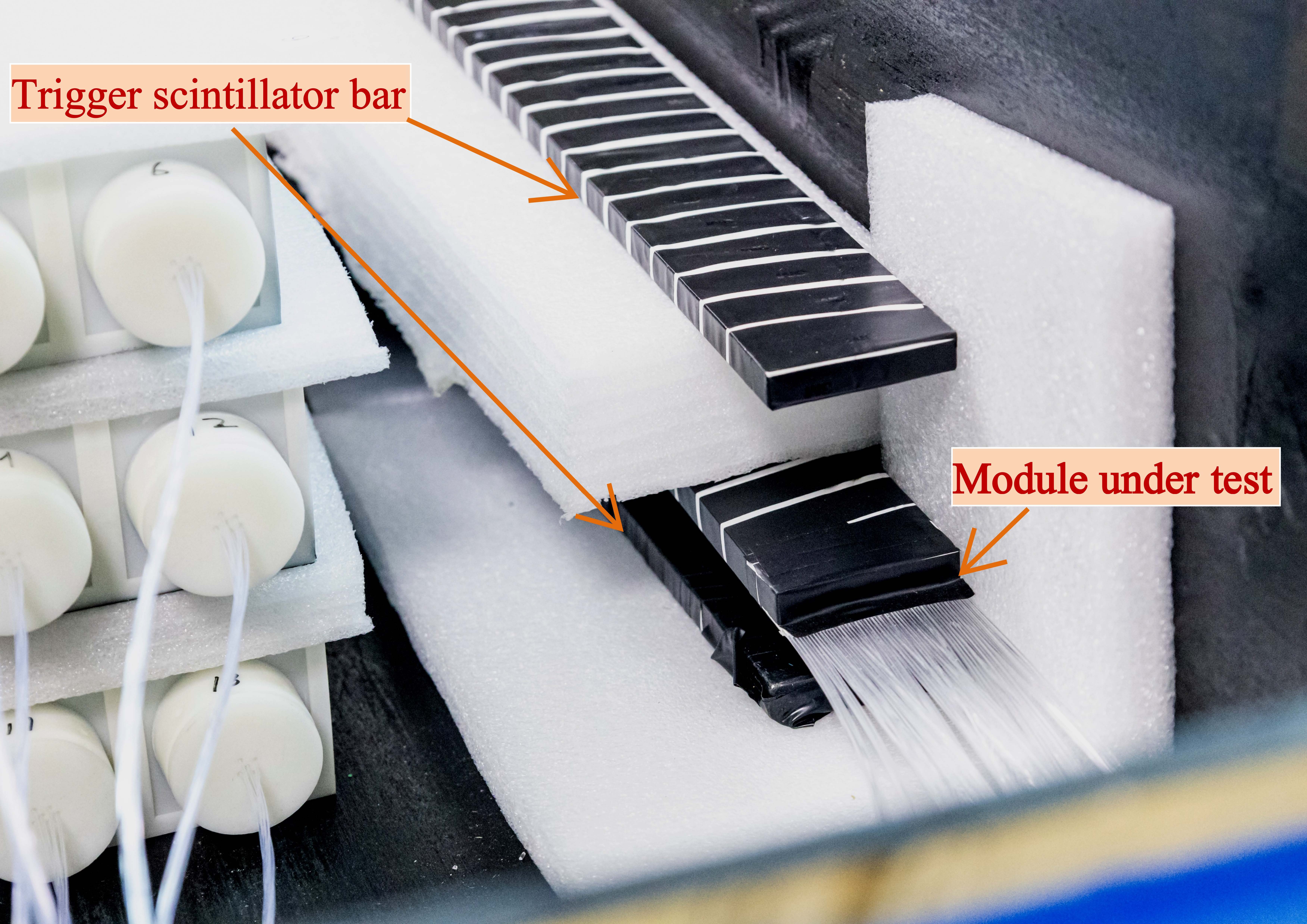}
  \caption{The single-module test setup. The module is positioned in
    the center, with scintillating bars of identical dimensions placed
    above and below to provide coincidence triggering.}
  \label{fig:TestSingleModule}
\end{figure}

The detection performance of each module is tested to guarantee
uniform response across the telescope.
A module under test is equipped with 19 detection channels, including
eighteen channels for fiber bundles and one channel for the
scintillating bar, and each channel is directly coupled to a PMT.
Waveforms are acquired for the module test, utilizing CAEN digitizers
V1743 (\SI{3.2}{GS/s}, 12-bit resolution) and DT5720 (\SI{250}{MS/s},
12-bit resolution). The former is used for the scintillator bar
channel and 14 fiber channels, while the latter is for other 4 fiber
channels.
Two identical scintillating bars (named trigger bars), each coupled to
a PMT, were installed on the module's top and bottom surfaces to form
a cosmic-ray muon trigger system, as shown in
\fig{TestSingleModule}.
The system generates an external trigger when signals from the trigger
bars are in coincidence.

\begin{figure}[!htb]
  \centering%
  \includegraphics[width=0.72\linewidth]{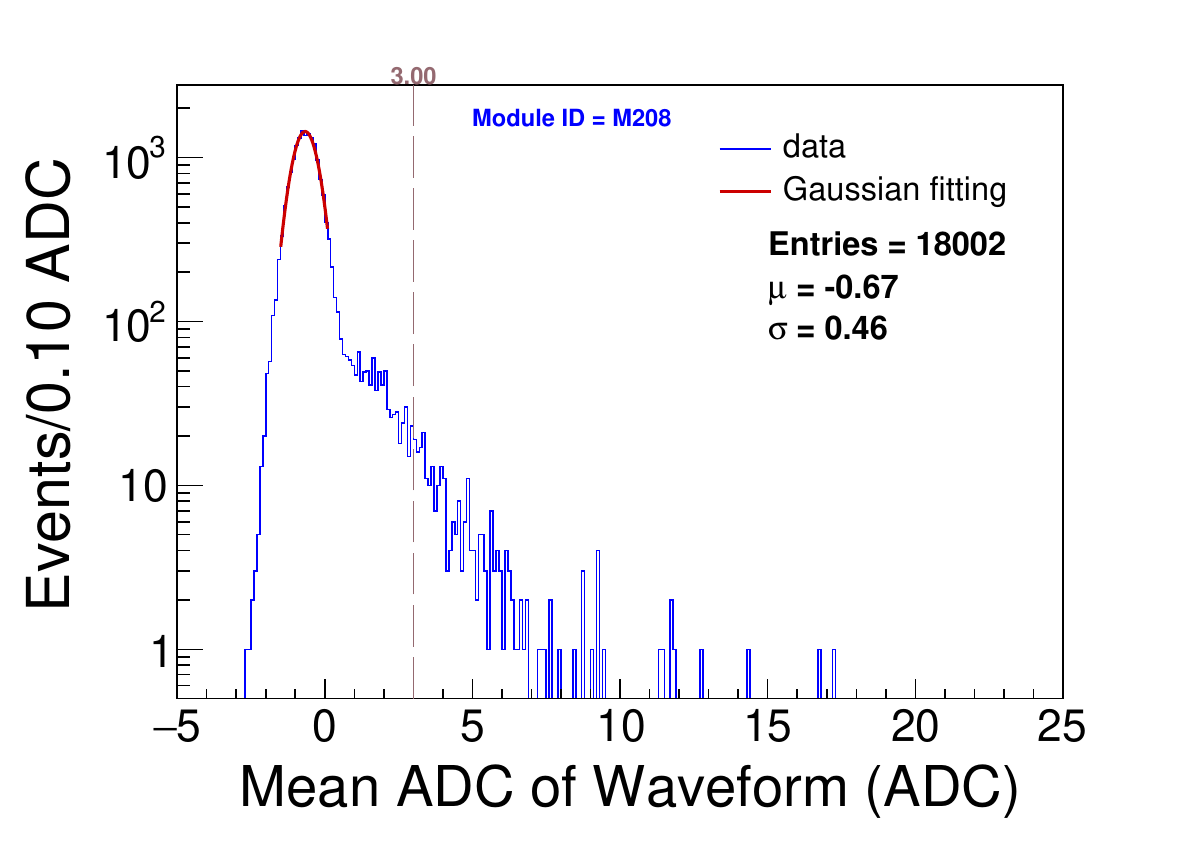}
  \caption{Distribution of the mean ADC in each waveform for a channel
    in the single module test. The pedestal is fitted with a Gaussian
    (red solid line), and the threshold, $\mu + 8\sigma$, is
    determined (brown dashed line).}
  \label{fig:waveformMeanADC}
\end{figure}

\begin{figure}[!htb]
  \centering%
  \includegraphics[width=0.72\linewidth]{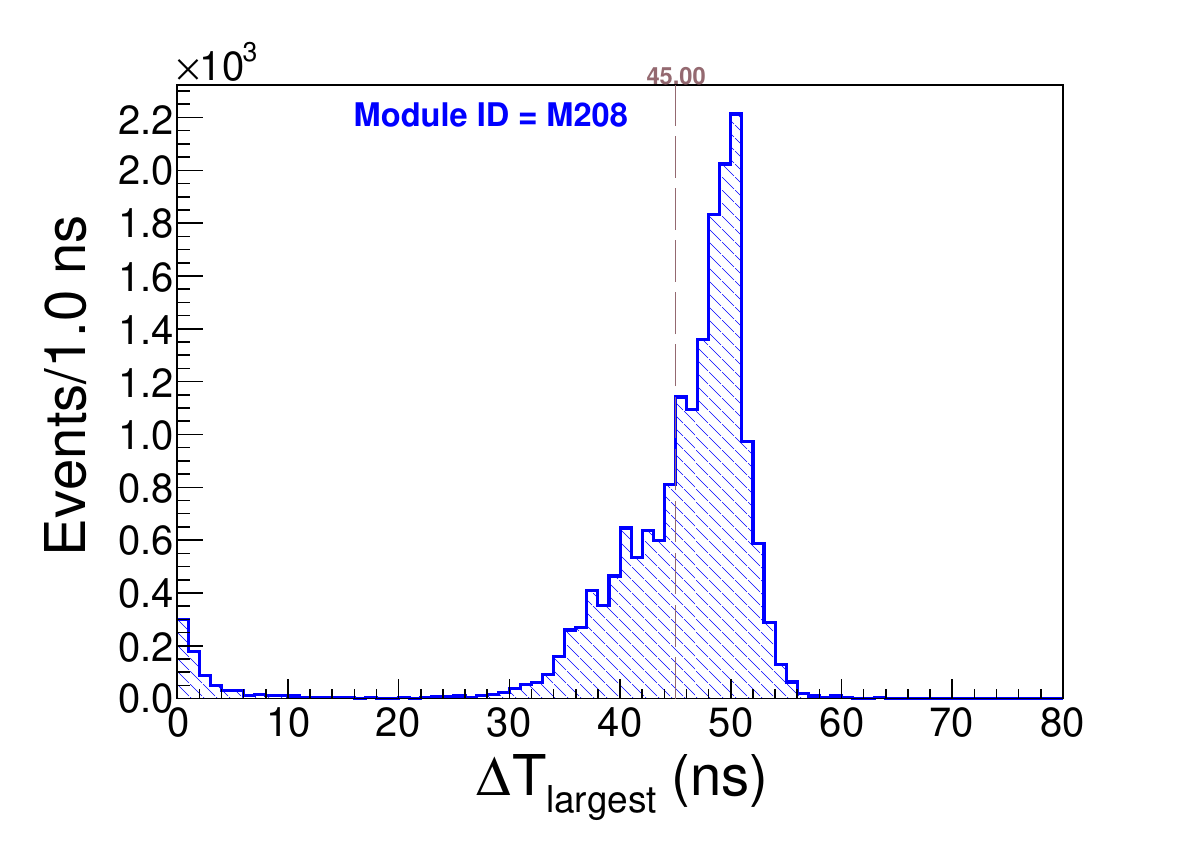}
  \caption{Distribution of the largest width above the threshold in
    each waveform for a channel in the single module test. See text
    for details.}
  \label{fig:waveformWidth}
\end{figure}

\begin{figure}[!htb]
  \centering%
  \includegraphics[width=0.72\linewidth]{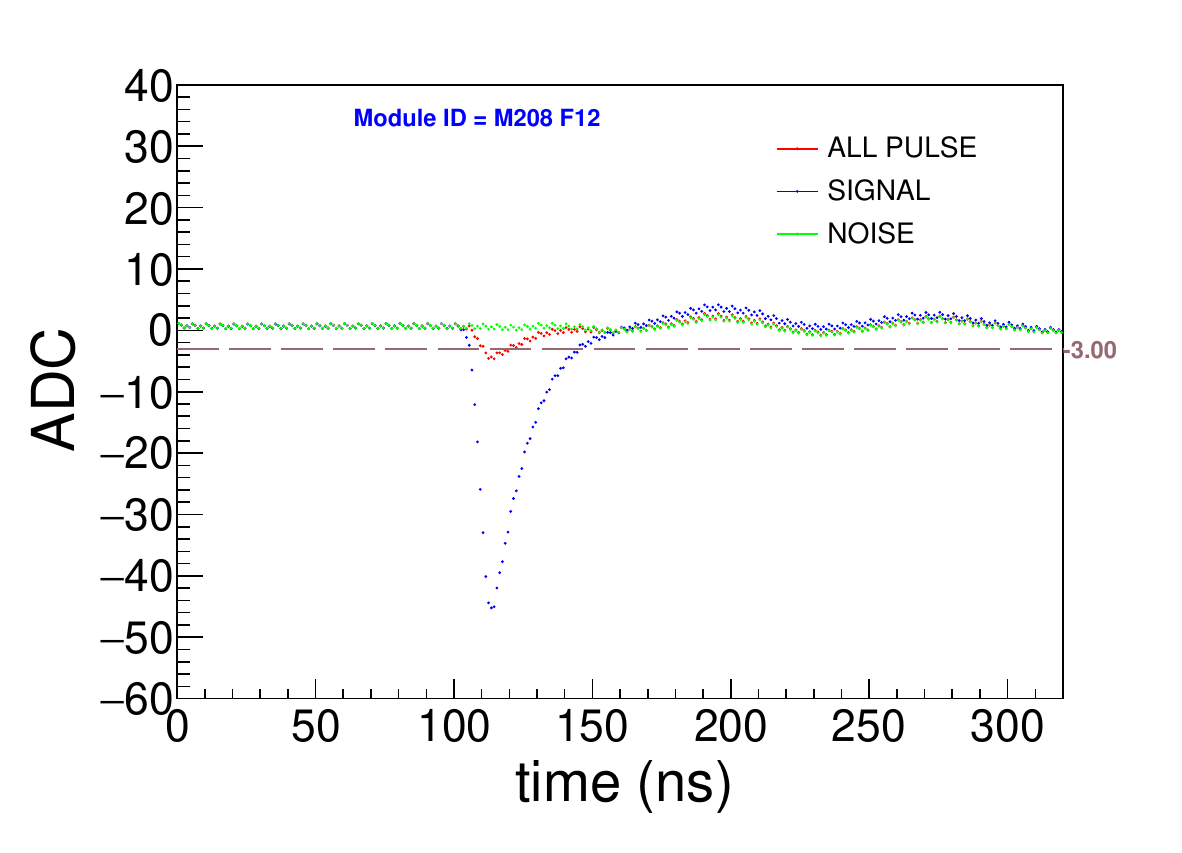}
  \caption{Profile distributions of waveforms for a fiber channel in
    the single module test.
    Each profile represents an accumulation of waveforms acquired, and
    each point in a profile is the average of ADCs with uncertainty at
    some time.
    With the determined selection criteria, waveforms (red) are
    separated as those of signals (blue) and noise (green). See
    text for details.
  }
  \label{fig:TestSingleModuleWaveDistribution}
\end{figure}

For each fiber and bar channel in the module,
the average ADC (reversed) of each waveform is filled into a
histogram, in which the pedestal is fitted with a Guassian, as shown
in \fig{waveformMeanADC}. The obtained mean ($\mu$) and uncertainty
($\sigma$) are then used to determine a threshold, $\mu + 8\sigma$,
for the channel.
The threshold is applied to each waveform, of which a few segments
above the threshold can be obtained. The time window of the widest
segment is filled into a histogram, as shown in \fig{waveformWidth},
and a conservative value, \SI{45}{ns}, of the waveform width for
signal is determined.
%
%
After applying the selection criteria, namely the threshold and the
waveform width, waveforms of signal and noise exhibit clear
separation, as shown in \fig{TestSingleModuleWaveDistribution}.

\begin{figure}[!htb]
  \centering%
  \includegraphics[width=0.72\linewidth]{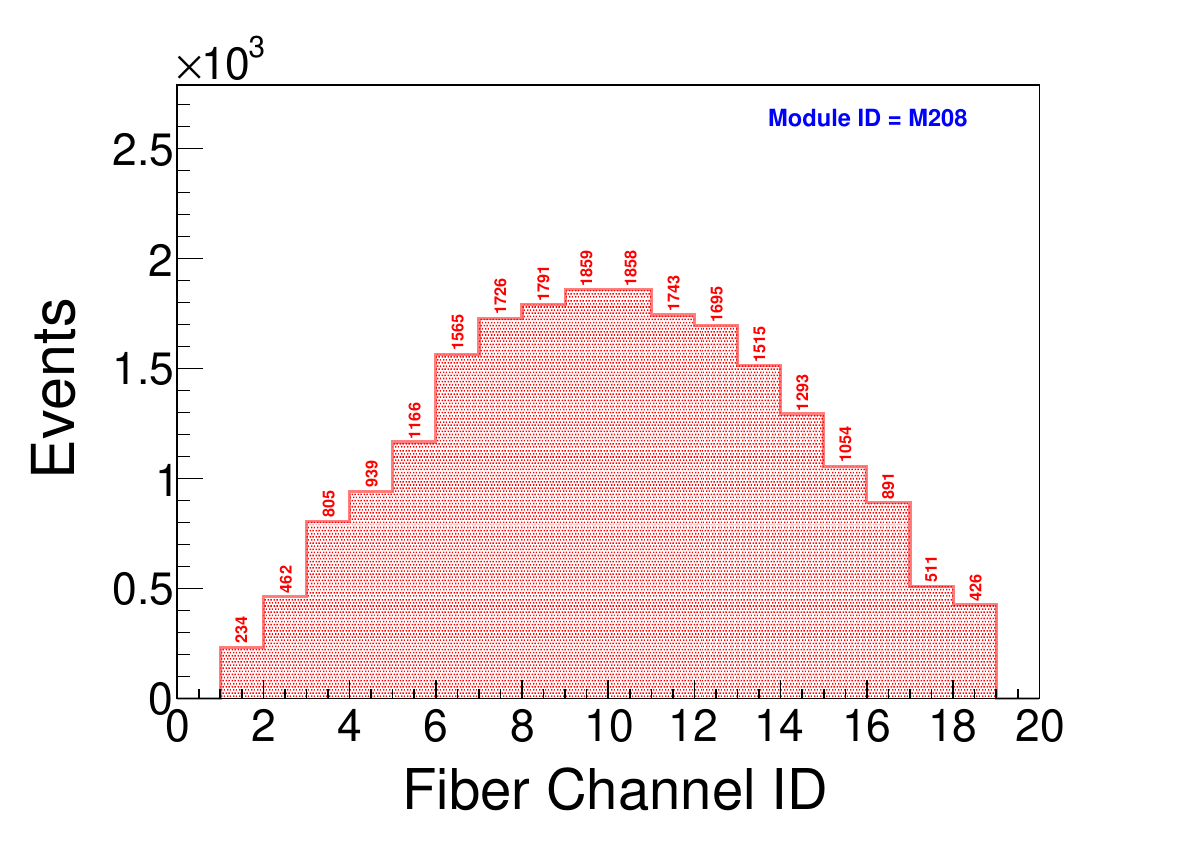}
  \caption{Distribution of fiber bundle response. (Illustrated using
    the eighth module in the second layer as an example)}
  \label{fig:TestSingleModuleHitMap}
\end{figure}

The analysis of responses from individual fiber bundle channels within
a single module, as shown in \fig{TestSingleModuleHitMap},
exhibits a uniform response in the central region and decreases
gradually toward the edges.
This distribution is consistent with expectations: the decreased
response at the edges arises from a smaller effective acceptance angle
and a corresponding decrease in particle hit probability.
All seventy-two modules have been tested, with all fiber channels
performing as expected.

\subsection{Uniformity Test of a Single Detection Layer}

\begin{figure}[!htb]
  \centering%
  \includegraphics[width=0.89\linewidth]{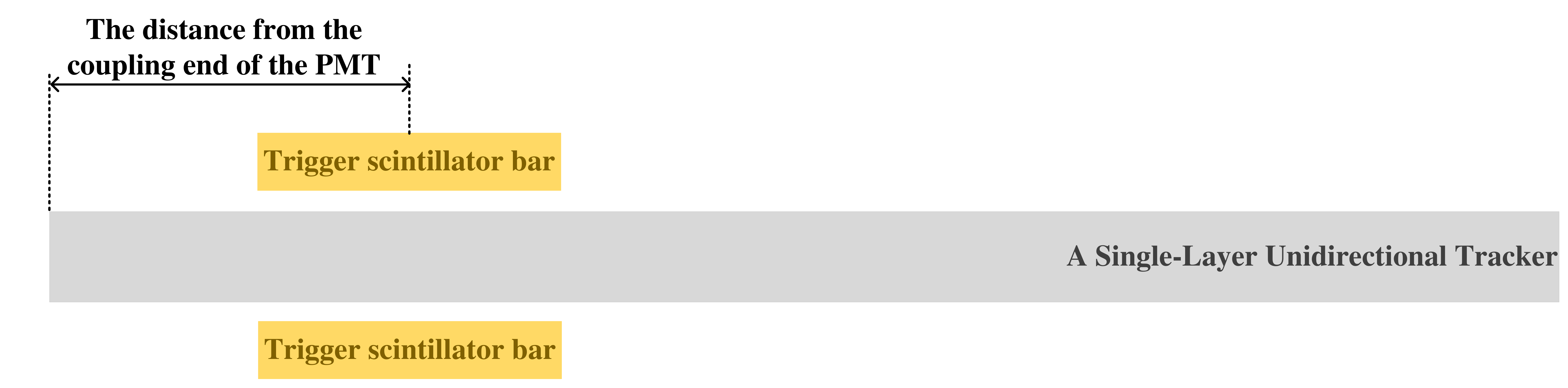}
  
  \bigskip
  \includegraphics[width=0.62\linewidth]{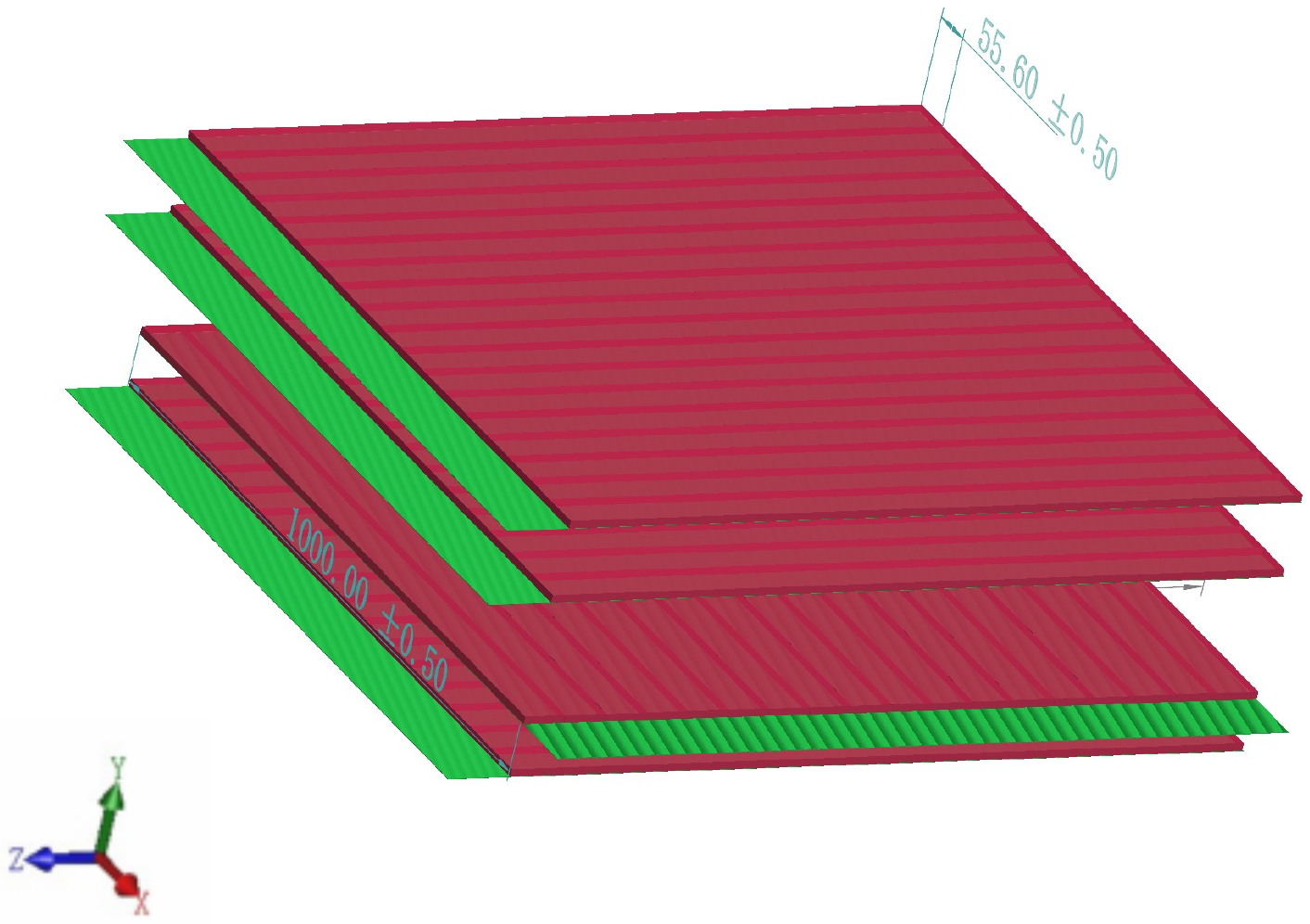}
  \caption{The experimental setup of uniformity test for scintillating
    bars (top) and fibres (bottom) in a detection layer.}
  \label{fig:Test_UniformitySetup}
\end{figure}

A dedicated experiment was carried out to evaluate the variation of
the detection efficiency with the distance between the hit position
and the PMT-coupled end of the scintillating bar, to assess the
uniformity of the detection layers.
The experiment utilized an external trigger mode with two identical
scintillator bars (\qtyproduct{250 x 100 x 1000}{\mm})
positioned above and below the detection layer, each coupled to a PMT
as a trigger source, as shown in \fig{Test_UniformitySetup} (top).

\begin{figure}[!htb]
  \centering%
  \includegraphics[width=0.72\linewidth]{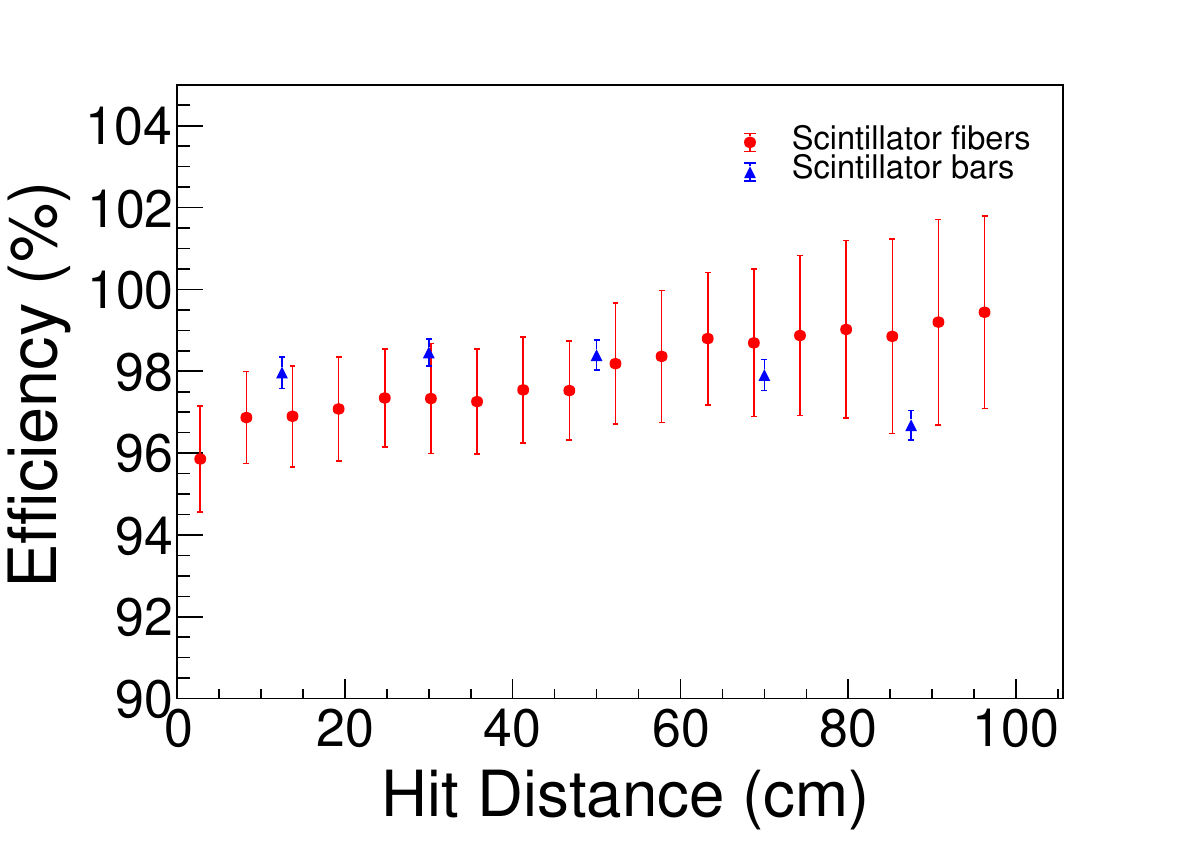}
  \caption{Detection efficiency of scintillating bars (blue) and
    fibers (red) in a detection layer as a function of hit distance
    with respect to the PMT-coupled end of the bar (see text for
    more explanations).}
  \label{fig:Test_UniformityResult}
\end{figure}

The trigger scintillating bars were positioned at \SI{12.5}{cm},
\SI{30}{cm}, \SI{50}{cm}, \SI{70}{cm}, and \SI{87.5}{cm} respectively
from the PMT-coupled end to quantify distance-dependent efficiency
variations.
The measured efficiency dependence on the trigger positions are shown
in \fig{Test_UniformityResult}.

Simultaneous analysis of the scintillating fiber signals was not
feasible under the external trigger mode due to the limited number of
DAQ channels.
Instead, three detection layers of the telescope were aligned in
parallel, while the remaining layer was arranged orthogonally to the
other three, as shown in \fig{Test_UniformitySetup} (bottom).
The modular design of the telescope allows the measurement of the
position dependence of detection efficiency. This approach provides a
direct evaluation of the fiber mats at different hit locations, as
shown in \fig{Test_UniformityResult}.

The measured detection efficiencies vary modestly across the
scintillating bars and fibers.
As the \emph{hit distance} in \fig{Test_UniformityResult} is refer to
the PMT-coupled end of a bar, which is opposite to that of fibers, the
two distributions show somehow opposite trends.
The variations are about 2\% for bars and 6\% for fibers, showing that
the layer-wise detection efficiencies are roughly uniform.

\section{Detector Performance}\label{sect:Resolution and efficiency}

\subsection{Hit Position Resolution}
\begin{figure}[!htb]
  \centering%
  \includegraphics[width=0.62\linewidth]{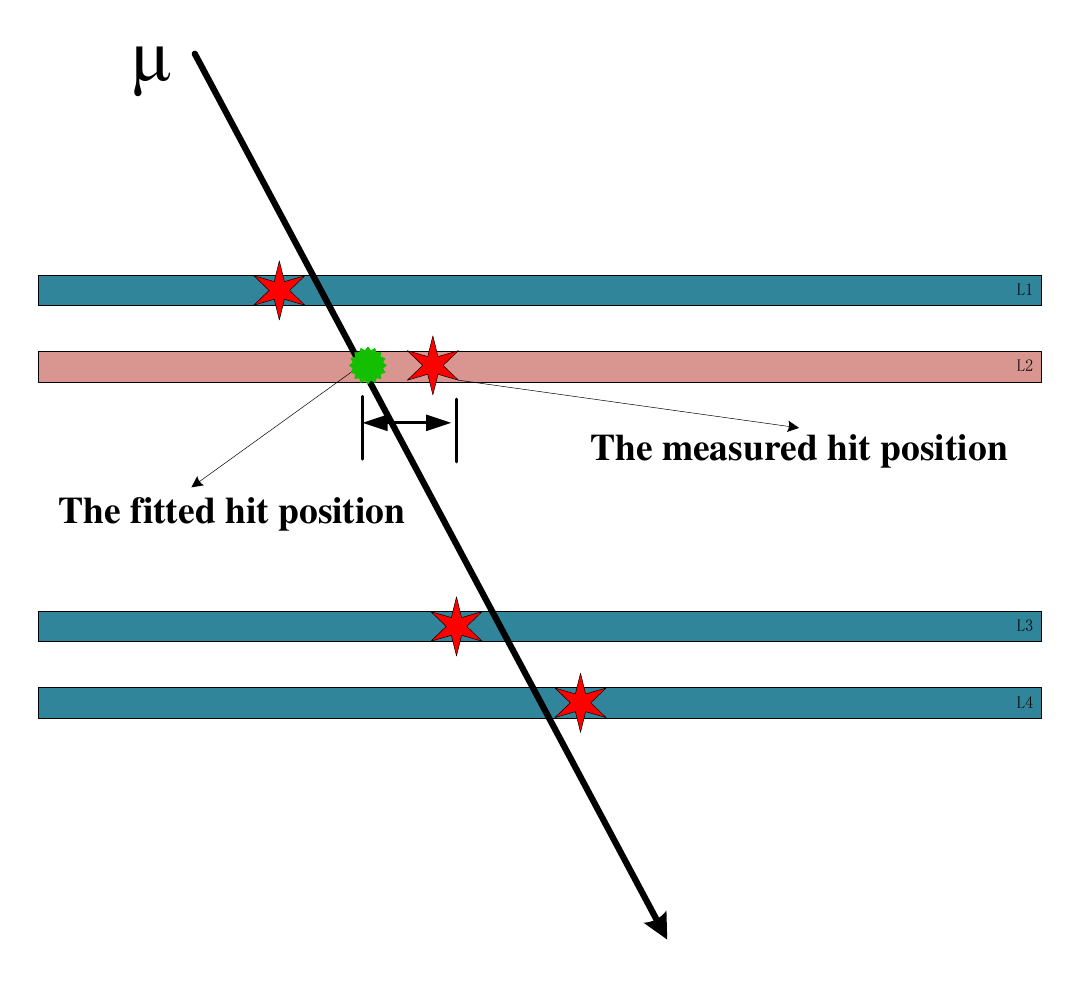}
  \caption{Schematic diagram of the spatial resolution test.}
  \label{fig:SpatialResolutionSetUp}
\end{figure}

\begin{figure}[!htb]
  \centering%
  \includegraphics[width=0.72\linewidth]{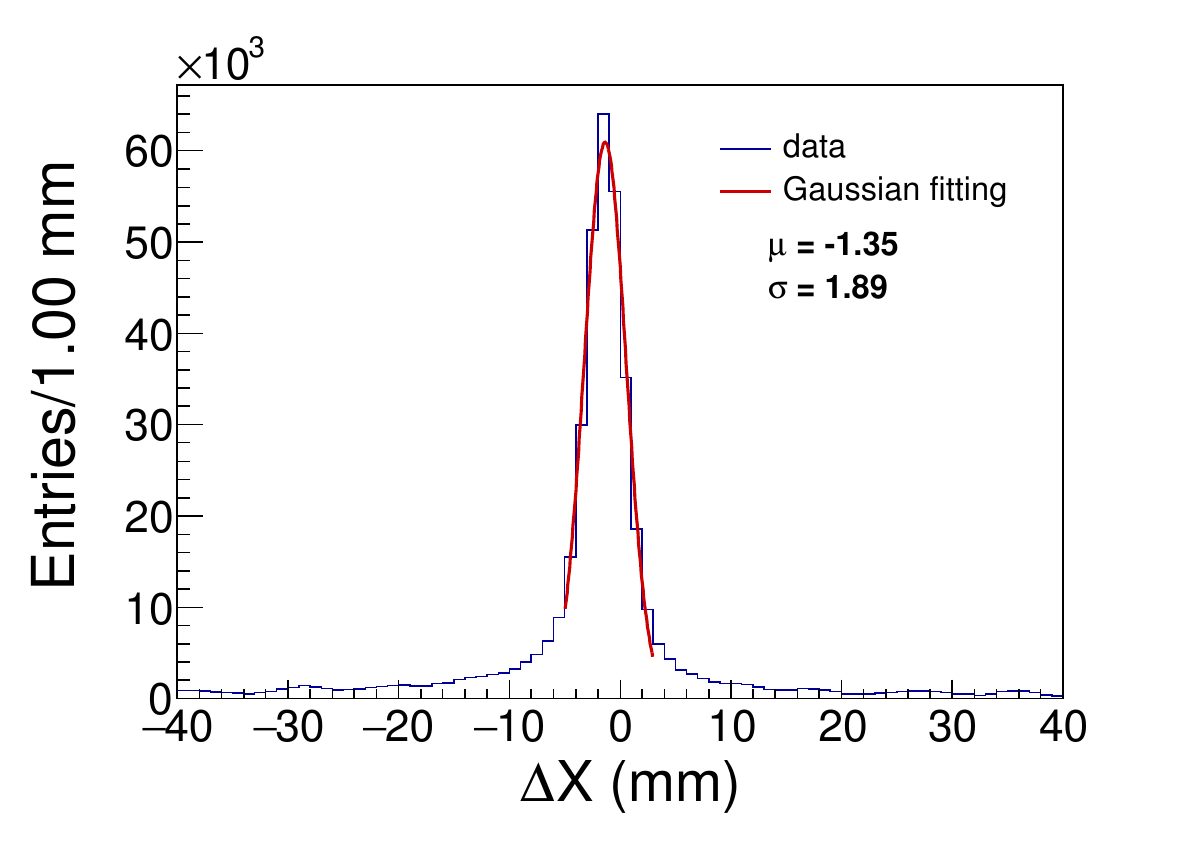}
  \caption{Residual distribution of the position measurement for a
    detection layer. The distribution is fitted with a Gaussian
    (red). See text for details.}
  \label{fig:SpatialResolutionResult}
\end{figure}

The telescope was self-calibrated by aligning the four detection
layers in parallel. Muon trajectories were reconstructed from three of
four layers to evaluate the spatial resolution of the remaining one.
Take the second layer as an example, the hit positions from the other
three layers were used to reconstruct the muon trajectory via the
least-squares method, yielding the position $x_{\rm rec}$.
The residual distribution was then derived by comparing $x_{\rm rec}$
with the actual observed position $x_{\rm obs}$ in the second layer,
defined as $\Delta x = x_{\rm obs} - x_{\rm rec}$, as shown in
\fig{SpatialResolutionSetUp}.
The residual distribution reflects both the spatial resolution of the
detector and possible mechanical alignment deviations.
The Gaussian standard deviation of the residual distribution gives a
measurement for spatial resolution, as shown in
\fig{SpatialResolutionResult} for the second layer. The results of all
four layers are listed in \tab{resolutions}.
%
%
Assuming the intrinsic resolutions of all layers are the same,
i.e. $\sigma_0$. For the straight line fit, the
variance of residual, $\sigma^2$, is given by (ignoring statistical
uncertainties),
\begin{equation*}
  \sigma^2 = \sigma_0^2 \left[ 1 + \frac{1}{L}
    + \frac{1}{L}\frac{(z_d-\xbar z)^2}{\xbar{z^2}-\xbar z^2} \right]
\end{equation*}
where $L$ is the number of layers excluding DUT, $z_d$ is the position
of DUT, and $z$ for other layers. With our test configuration, the
intrinsic resolution can be obtained as $ 1.45\pm 0.12\mm$.

\begin{table}[!htb]
  \centering
  \caption{Residual resolutions measured for all four detection layers.}
  \label{tab:resolutions}
  \begin{tabular}{ccccc}
    \hline\hline
    Detection Layer &1&2&3&4
    \\
    $z$ position (mm)
    & 0.0 & 136.8 & 400.2 & 533.0
    \\
    Resolution (\si{\um})
    & $2.172$ & $1.893$ & $1.867$ & $2.492$
    \\\hline
  \end{tabular}
\end{table}

\subsection{Detection efficiency}

The detection efficiency of the telescope is mainly decided by that of
scintillating fibers.
As for a detection layer, the efficiency of the fiber layer is defined
as $\epsilon = N_f/N_b$, where $N_b$ is the total number of tracks
passing through a given scintillating bar (module), and $N_f$ is the
number of tracks having at least one hit among all fiber channels.
The reason for this definition is that the hit position of fibers can
not be determined by themselves alone.

\begin{figure}[!htb]
  \centering%
  \includegraphics[width=0.72\linewidth]{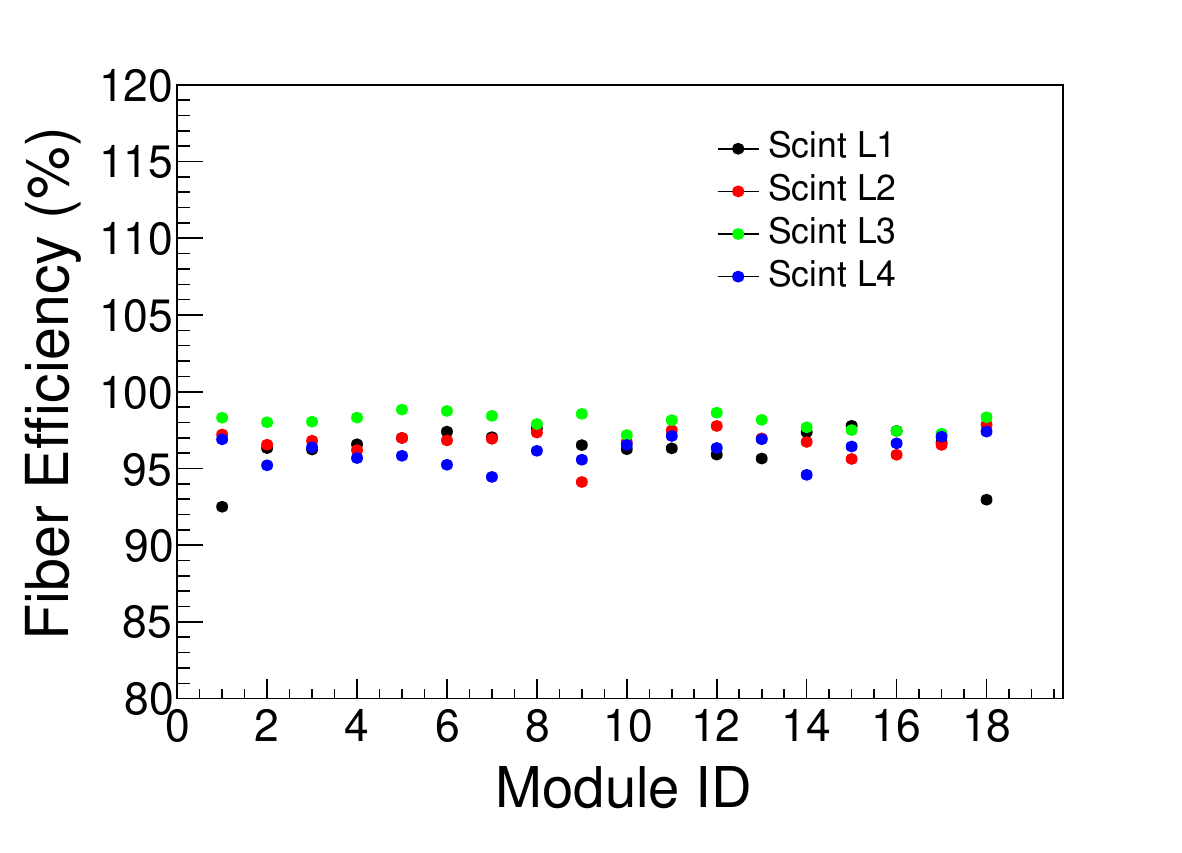}
  \caption{Detection efficiency of the fiber layer as a function of
    module for four layers (different colors).  }
  \label{fig:FiberEffDistributionResult}
\end{figure}

The efficiency variation with modules for all layers are shown in 
\fig{FiberEffDistributionResult}.
The results show that the detection efficiency of a single (fiber)
layer exceeds approximately 96\%.
The overall efficiency of the telescope is estimated as the product of
those of four layers, giving 85\%.

\section{Summary and outlook}\label{sect:Summary}
We have developed a cosmic-ray muon telescope featuring a novel
position reconstruction method based on encoding signals from
scintillating bars and fibers.
This approach achieves millimeter-scale spatial resolution, with high
detection efficiency, while minimizing the number of readout
electronics channels.
The optimized encoding scheme enhances flexibility and scalability,
enabling cost-effective experimental applications.

Experimental results demonstrate that the telescope achieves a spatial
resolution better than \SI{2}{mm} and a detection efficiency above
85\%.
These results validate the encoding strategy for precise and efficient
particle tracking.
The telescope is currently utilized in the ground calibration of the
HERD CALO.
With little modifications, the design could be adapted for other kind
of experiments, maintaining performance while controlling costs.

\section*{Acknowledgements}
This work was supported by the National Key R\&D program of China
(2021YFA0718403) and Shandong University Undergraduate Education
Research Project (2024Y223).

\bibliographystyle{unsrt2authabbrvpp}
\bibliography{niuy.bib}

\end{document}